%% file: paper.tex
\documentclass[submission, Phys]{SciPost}

\input{incl_settings.tex} 
\input{incl_shortcuts.tex}

\graphicspath{{./figs/}}

\begin{document}

% \vspace*{-2.5em}
% \hfill{\small IRMP-CP3-22-56, MCNET-22-22, FERMILAB-PUB-22-915-T}
% \vspace*{0.5em}

\begin{center}{\Large \textbf{
The Landscape of Unfolding with Machine Learning 
}}\end{center}

% authors ordered by those who wrote new code and produced new results for the project and those who collected, commented, helped, etc...
\begin{center}
Nathan Huetsch\textsuperscript{1},
Javier Mariño Villadamigo\textsuperscript{1},
Alexander Shmakov\textsuperscript{2}, 
Sascha Diefenbacher\textsuperscript{3}, \\
Vinicius Mikuni\textsuperscript{3}, 
Theo Heimel\textsuperscript{1},
Michael Fenton\textsuperscript{2},
Kevin Greif\textsuperscript{2},\\
Benjamin Nachman\textsuperscript{3,4},
Daniel Whiteson\textsuperscript{2},
Anja Butter\textsuperscript{1,5},
and Tilman Plehn\textsuperscript{1,6}
\end{center}

\begin{center}
{\bf 1} Institut für Theoretische Physik, Universität Heidelberg, Germany
\\
{\bf 2}  Department of Physics and Astronomy, University of California, Irvine, USA
\\
{\bf 3} Physics Division, Lawrence Berkeley National Laboratory, Berkeley, USA
\\
{\bf 4} Berkeley Institute for Data Science, University of California, Berkeley, USA
\\
{\bf 5} LPNHE, Sorbonne Universit\'e, Universit\'e Paris Cit\'e, CNRS/IN2P3, Paris, France
\\
{\bf 6} Interdisciplinary Center for Scientific Computing (IWR), Universit\"at Heidelberg, Germany
\end{center}

\begin{center}
\today
\end{center}

% For convenience during refereeing: line numbers
%\linenumbers

\section*{Abstract}
         {\bf Recent innovations from machine learning allow for data unfolding, without binning and including 
           correlations across many dimensions. We describe a set of 
           known, upgraded, and new methods for ML-based unfolding.  The performance of these approaches are evaluated on the same two datasets. We find that all techniques are capable of accurately reproducing the particle-level spectra across complex observables.  Given that these approaches are conceptually diverse, they offer an exciting toolkit for a new class of measurements that can probe the Standard Model with an unprecedented level of detail and may enable sensitivity to new phenomena.}

\clearpage
\vspace{10pt}
\noindent\rule{\textwidth}{1pt}
\tableofcontents\thispagestyle{fancy}
\noindent\rule{\textwidth}{1pt}
\vspace{10pt}

\clearpage
%%%%%%%%%%%%%%%%%%%%%%%%%%%%%%%%%%%%%%%%%%%%%%%%%%%
\section{Introduction}
\label{sec:intro}

Particle physics experiments seek to reveal clues about the fundamental properties of particles
and their interactions.  A key challenge is that predictions from quantum 
field theory are at the level of partons, while experiments observe the corresponding
detector signatures.
Precise and detailed simulations link these two levels~\cite{Campbell:2022qmc}. 
They fold predictions for the hard process through QCD effects, hadronization, and the 
detector response to compare with data.  This statistically powerful {\it forward inferences} approach has been widely used.  

However, forward inference requires access to the data and the detector simulation. These conditions are rarely satisfied outside of a given 
experiment, severely limiting the ability of the broader community to study particle 
physics data.
In addition, analysis of data from the high-luminosity LHC with forward inference will require precise simulations for every hypothesis,  challenging available computing resources.

An alternative approach is {\it unfolding}. Rather than correcting predictions for the effects of the detector, the data are 
adjusted to provide an estimate of their pre-detector distributions.  Since the effects described 
by our forward simulation are stochastic, this adjustment is performed on a statistical basis.
Unfolding offers important advantages, such as making data analysis possible by a broader 
community and enabling an efficient combination of data from several experiments, such as in 
global analyses of the Standard Model Effective Theory~\cite{Brivio:2019ius,Elmer:2023wtr}.

Traditional unfolding algorithms have been used extensively, successfully delivering a multitude of 
differential cross section measurements~\cite{Cowan:2002in,Spano:2013nca,Arratia:2021otl}.
The most widely-used methods are Iterative Bayesian
Unfolding~\cite{1974AJ.....79..745L,Richardson:72,Lucy:1974yx,DAgostini:1994fjx},
Singular Value Decomposition~\cite{Hocker:1995kb}, and
TUnfold~\cite{Schmitt:2012kp}.  However, each of these methods can only be applied to binned datasets of small 
dimensionality, such that the unfolded observables and their binning have to be selected in advance. 

Machine learning (ML) techniques have revolutionized unfolding by allowing for unbinned cross 
sections to be measured across many dimensions~\cite{Arratia:2021otl,Butter:2022rso}.  Where 
sufficient information is unfolded, new observables can be calculated from unbinned data, long after 
the initial publication.
The first ML-based unfolding method applied to data is OmniFold~\cite{Andreassen:2019cjw,Andreassen:2021zzk}, which uses classifiers to reweight simulations. 
It has recently been applied to studies of hadronic final states at
H1~\cite{H1:2021wkz,H1prelim-22-031,H1:2023fzk,H1prelim-21-031},
LHCb~\cite{LHCb:2022rky}, CMS~\cite{Komiske:2022vxg}, and
STAR~\cite{Song:2023sxb}.  Alternative ML-unfolding methods use generative networks, 
either for distribution mapping~\cite{Datta:2018mwd,Howard:2021pos,Diefenbacher:2023wec,Butter:2023ira}
or for probabilistic, conditional generation~\cite{Bellagente:2019uyp,Bellagente:2020piv,Vandegar:2020yvw,Backes:2022vmn,Leigh:2022lpn,Ackerschott:2023nax,Shmakov:2023kjj,Shmakov:2024xxx}. 

The goal of this paper is to lay out and extend the landscape of ML methods.
We benchmark a diverse 
set of approaches on the same datasets, to facilitate direct comparisons.  Some methods have been studied with an 
iterative component to mitigate the sensitivity to starting particle-level simulations.
To simplify the setup and reduce stochastic effects from iterating, we apply all methods 
with only a single step. The goal is to estimate the posterior with the starting 
simulation as the prior. Performing this step well is the essential component of a full unfolding 
approach.

We begin with a brief introduction 
of the different methods for ML-based unfolding in Sec.~\ref{sec:ml}.
In Sec.~\ref{sec:omni}, we show how all approaches can accurately unfold from detector level to the particle level using a $Z$+jets benchmark dataset.
For certain theory questions it is useful to further unfold 
to the parton level, treating QCD radiation as a distortion to be corrected like detector effects.  
As an example of this type of unfolding, we study top quark pair production in Sec.~\ref{sec:tops}. 
In Sec.~\ref{sec:outlook} we 
summarize the advantages of the different methods, to help
the experimental collaborations pick the method(s) best-suited 
for a given task. The figures shown in App.~\ref{sec:combined} just 
combine results from the $Z$+jets study in Sec.~\ref{sec:omni}.

%%%%%%%%%%%%%%%%%%%%%%%%%%%%%%%%%%%%%%%%%%%%%%%%%%%
\section{ML-Unfolding}
\label{sec:ml}

We define our unfolding problem using four phase space densities, which are encoded in the corresponding samples,
in the sense of unsupervised density estimation in ML-terms.
We rely on simulated predictions at the
particle/parton level, $\psimp(\xp)$, and the detector or reconstruction (reco) 
level, $\psimr(\xr)$.
Unfolding turns the measured $\pd$ into
$p_\text{unfold}$,
\begin{alignat}{9}
  & \psimp
  \quad \xleftrightarrow{\text{unfolding inference}} \quad 
  && p_\text{unfold}(\xp)
  \notag \\
  & \hspace*{-9mm} \text{\footnotesize simulation} \Bigg\downarrow
  && \hspace*{+6mm} \Bigg\uparrow \text{\footnotesize unfolding}
  \notag \\
  & \psimr 
  \quad \xleftrightarrow{\text{\; forward inference \;}} \quad 
  && \pd(\xr)
\label{eq:schematic}
\end{alignat}
Our simulated samples come in pairs
$(\xp,\xr)$, which can be used for unfolding. Data only exist on the $\xr$ level. The
features of the unfolded data $p_\text{unfold}$ should be determined
by $\pd$, but will always include a data-independent 
bias 
from the assumed $\psimp$. The question how we can minimize
the resulting model dependence will be part of a follow-up of this
study.

Established ML-techniques for unfolding rely on one of two approaches. They
either reweight simulated samples, or they generate
unfolded samples from conditional probabilities. We will briefly
introduce both original
methods~\cite{Andreassen:2019cjw,Bellagente:2019uyp,Bellagente:2020piv},
as well as a more recent hybrid approach of mapping distributions
using generative networks.

%%%%%%%%%%%%%%%%%%%%%%%%%%%%%%%%%%%%%%%%%%%%%%%%%%%
\subsection{Reweighting: (b)OmniFold}
\label{sec:ml_omni}

The deep learning-based approach to unfolding via re-weighting is
OmniFold~\cite{Andreassen:2019cjw,Andreassen:2021zzk}. It is based on the Neyman--Pearson
lemma~\cite{neyman1933ix}, stating that an optimally trained, calibrated classifier $C$
will learn the likelihood ratio of the two underlying phase space
distributions. If we use a binary cross entropy (BCE) loss to distinguish between data and simulated
reco-level events, then the following combination approximates the likelihood ratio:
\begin{align}
%  C(\xr) = \frac{ \sim(\xr)}{ \psimr+ \pd(\xr)}
%  \qquad \Leftrightarrow \qquad 
  w(\xr) \equiv \frac{\pd(\xr)}{\psimr(\xr)} = \frac{C(\xr)}{1-C(\xr)} \; .
\label{eq:likelihood_ratio}
\end{align}
OmniFold computes these classifier weights at the reco-level, and uses the paired simulated data to pull these weights from the reco-level
events to the particle-level events. The re-weighted simulated events
then define
\begin{align}
  \punf(\xp) = w(\xr) \; \psimp (\xp) \; .
\end{align}
This weight-pushing is the first step in the two-step iterative OmniFold
algorithm. Because we are leaving out the
model dependence to a dedicated second study, we restrict ourselves to
this first iteration, which in the scheme of 
Eq.\eqref{eq:schematic} looks like
\begin{alignat}{9}
  & \psimp
  \quad \xrightarrow{\text{\; classifier weights (3) \;}} \quad 
  && \punf(\xp)
  \notag \\
  & \hspace*{-16mm} \text{\footnotesize pull (2)/push weights (4)} \Bigg\updownarrow
  && %\hspace*{+6mm} \Bigg\uparrow \text{\footnotesize reweighting}
  \notag \\
  & \psimr 
  \quad \; \xleftrightarrow{\text{\; classifier weights (1) \;}} \quad 
  && \pd(\xr)
\label{eq:schematic_omni}
\end{alignat}

\paragraph{Bayesian network}
Bayesian versions can be derived for any deterministic neural network
with a likelihood
loss~\cite{bnn_early,bnn_early2,bnn_early3,kendall_gal,Plehn:2022ftl}. The
BNN training does not fix the network parameters, but allows them to
learn distributions, such that sampling over the network parameters
gives the probability distribution in model space, \ie for the network
output. Based on studies for
regression~\cite{Bollweg:2019skg,Badger:2022hwf} and classification
tasks~\cite{Kasieczka:2020vlh}, there is evidence that for a sufficiently
deep network we can assign independent Gaussians to each network
parameter~\cite{bnn_early3}. This effectively doubles the size of the network which now
learns a central prediction and the error bar simultaneously.  Even though the weights are Gaussian distributed, the final network output is generally not a Gaussian. As we will see below, 
Bayesian networks can be generalized to generative tasks~\cite{Bellagente:2021yyh,Butter:2021csz,Butter:2023fov}.

One benefit of Bayesian networks is that they automatically include
a generalized dropout and a weight regularization~\cite{molchanov2017variational,fortuin2022priors,Plehn:2022ftl}, derived from
Bayes' theorem together with the likelihood loss. This means that BNNs
are automatically protected from overtraining and 
an attractive option for applications where the precision of the network is
critical, like the classifier reweighting in OmniFold.

%%%%%%%%%%%%%%%%%%%%%%%%%%%%%%%%%%%%%%%%%%%%%%%%%%%
\subsection{Mapping distributions: Schr\"odinger Bridge and Direct Diffusion}
\label{sec:ml_distri}

Instead of reweighting phase-space events, we can use generative
neural networks to morph a base distribution to a target
distribution. In our case, we train a network to map event
distributions from $\xr$ to $\xp$ based on the paired or unpaired
simulated events and apply this mapping to $\pd(\xr)$ to
generate $\punf(\xp)$:
\begin{alignat}{9}
  & \psimp
%  \quad \xleftrightarrow{\text{unfolding inference}} \quad 
  && p_\text{unfold}(\xp)
  \notag \\
  & \hspace*{-6mm} \text{\footnotesize training} \Bigg\uparrow
  && \hspace*{+6mm} \Bigg\uparrow \text{\footnotesize distribution mapping}
  \notag \\
  & \psimr 
  \quad \xleftrightarrow{\text{\; correspondence \;}} \quad 
  && \pd(\xr)
\label{eq:schematic_didi}
\end{alignat}
As mentioned above, the trained mapping assumes that $\psimr$ and $\pd$
describe the same features at the reco-level.
Two ML-methods that we study for this task include
Schr\"odinger Bridges~\cite{Diefenbacher:2023wec} and Direct
Diffusion~\cite{Butter:2023ira}, see also Ref.~\cite{Datta:2018mwd} for an early study.  

%%%%%%%%%%%%%%%%%%%%%%%%%%%%%%%%%%%%%%%%%%%%%%%%%%%
\subsubsection{Schr\"odinger Bridge}
\label{sec:SB}

Schr\"odinger Bridges define the transformation between particle-level
events $\xp \sim \psimp$ to reco-level events $\xr \sim \psimr$ as a
time-dependent process following a forward-time stochastic
differential equation (SDE)
\begin{align}
    dx = f(x,t)\mathrm{d}t + g(t)dw \;.
    \label{eq:sde}
\end{align}
The drift term $f$ controls the deterministic part of the
time-evolution, $g$ is the noise schedule, and $dw$ a noise
infinitesimal. For such an SDE, the reverse time evolution follows the
SDE
\begin{align}
    dx = [f(x,t)-g(t)^2\nabla\log p(x,t)]dt + g(t)dw \; ,
    \label{eq:rsde}
\end{align}
with the corresponding score $s(x,t)=\nabla\log p(x,t)$. To construct an
unfolding, we need to find $f$ and $g$ for our forward process from
particle level to reco level, and then encode $s_\theta(x,t)$ in the
unfolding network~\cite{song2021scorebased}.

Constructing a forward-time SDE that transforms an arbitrary distribution into another 
is much more challenging than mapping a distribution into a noise distribution 
with known probability density (e.g. a Gaussian), as is the case for standard SDE-based diffusion networks.  
A framework to construct a transport plan in the general case was
proposed by Erwin Schr\"odinger~\cite{schrodinger1931umkehrung}. It
introduces two wave functions describing the 
time-dependent density as
$p(x,t)=\widehat{\Psi}(x,t)\Psi(x,t)$. By setting the drift coefficient to
$f=g(t)^2\nabla\log {\Psi}(x, t)$ the forward and reverse SDEs in Eqs.\eqref{eq:sde} and~\eqref{eq:rsde} become
\begin{align}
        dx &= \phantom{-} g(t)^2\nabla\log {\Psi}(x, t)dt  + g(t)dw 
        %        \label{eq:sb_sde}
        \notag \\ 
        dx &= - g(t)^2\nabla\log \widehat{\Psi}(x, t)dt  + g(t)dw \; .
        \label{eq:sb_sde}
\end{align}
If the two wave-functions fulfill the coupled partial differential equations
\begin{align}
  \frac{\partial{\Psi(x,t)}}{\partial t}
  &=  - \frac{1}{2} g(t)^2 \Delta \Psi(x,t) \notag \\
  \frac{\partial{\widehat{\Psi}(x,t)}}{\partial t}
  &= \phantom{-} \frac{1}{2} g(t)^2 \Delta \widehat{\Psi}(x,t) \; ,
    \label{eq:sb_pde}
\end{align}
with the boundary conditions
\begin{align}
  \widehat{\Psi}(x,t)\Psi(x,t)
  = \begin{cases}
    \psimp (x) \qqquad & t=0 \\
    \psimr (x) \qqquad & t=1  \; ,
  \end{cases}
\end{align}
then the SDEs in Eq.\eqref{eq:sb_sde} transform particle-level events
to reco-level events and vice versa.

Next, we need to find $\Psi$, $\widehat{\Psi}$ that fulfill the
conditions. The authors of Ref.~\cite{liu2023i2sb} observe that
reverse generation following Eq.\eqref{eq:sb_sde} does not require
access to the wave functions, but only to the score function
$\nabla\log \widehat{\Psi}$. For paired training data,
\begin{align}
  (x_0, x_1)\sim (\psimp, \psimr)
\end{align}
the posterior encoded in the SDEs in Eq.\eqref{eq:sb_sde}, conditioned
on the respective initial and final states, has the analytic form
\begin{align}
  q(x|x_0, x_1) &= \mathcal{N}(x_t;\mu_t(x_0,x_1),\Sigma_t) \notag \\
  \text{with} \qquad 
  \mu_t &= \frac{\bar{\sigma}_t^2}{\bar{\sigma}_t^2 + \sigma^2_t} x_0 +
  \frac{\sigma^2_t    }{\bar{\sigma}_t^2 + \sigma^2_t} x_1
  \quad \text{and} \quad
  \Sigma_t = \frac{\sigma_t^2 \bar{\sigma}_t^2}{\bar{\sigma}_t^2 + \sigma^2_t} \; ,
  \label{eq:sb_posterior}
\end{align}
denoting $\sigma^2_t = \int_0^t g^2(\tau) d\tau$ and $\bar{\sigma}^2_t
= \int_t^1 g^2(\tau) d\tau$. This allows for the generation of samples from this
stochastic process as $x_t(x_0, x_1) = \mu_t + \Sigma_t \epsilon$ with
$\epsilon \sim \mathcal{N}(0,1)$ and $(x_0, x_1)$, a pair of reco-level
and particle-level events. Moreover, the score $\nabla\log
\widehat{\Psi}$ can be learned by minimizing the loss
\begin{align}
  \loss_\text{SB}
  =  \XXLangle
  \left[ \epsilon_\theta(x_t(x_0,x_1),t) -  \frac{x_t(x_0,x_1) - x_0}{\sigma_t} \right]^2
  \XXRangle_{t \sim \mathcal{U}([0,1]),(x_0, x_1) \sim p(\xp, \xr)} \; ,
  \label{eq:loss_sb}
\end{align}
where $x_t$ is sampled according to Eq.\eqref{eq:sb_posterior}.
After training, the network unfolds by numerically solving the reverse
SDE Eq.\eqref{eq:sb_sde} with the $\xr$ values as the initial conditions.

We follow a slight variation, where the dynamics are reduced to a
deterministic process~\cite{liu2023i2sb}. This can be achieved by
replacing the posterior distribution Eq.\eqref{eq:sb_posterior} by its
mean and training the network to encode not the score function, but
the velocity field of the reverse process, which then takes the
form of an ordinary differential equation:
\begin{align}
    dx_t = v_t(x_t|x_0)dt = \frac{\beta_t}{\sigma_t^2}(x_t - x_0)dt \; .
\end{align}
For the noise schedule, we follow Ref.~\cite{Diefenbacher:2023wec} and
use $g(t)=\sqrt{\beta(t)}$, with $\beta(t)$ the triangular function
\begin{align}
    \beta(t) = \begin{cases}
    \beta_{0} + 2 (\beta_{1} - \beta_{0}) t & 0 \leq t < \frac{1}{2} \\
    \beta_{1} - 2 (\beta_{1} - \beta_{0}) \left( t - \frac{1}{2}\right) & \frac{1}{2} \leq t \leq 1  \; .
\end{cases}
\end{align}
with $\beta_0 = 10^{-5}$ and $\beta_1 = 10^{-4}$.

%%%%%%%%%%%%%%%%%%%%%%%%%%%%%%%%%%%%%%%%%%%%%%%%%%%
\subsubsection{Direct Diffusion}
\label{sec:diff}

Like the Schr\"odinger Bridge, Direct Diffusion (DiDi)
describes a time evolution between particle-level 
events at $t=0$ and reco-level events at $t=1$. Following 
the Conditional Flow
Matching (CFM)~\cite{lipman2023flow} framework, DiDi uses an
ordinary differential equation (ODE)
\begin{align}
    \frac{dx(t)}{dt} = v_\theta(x(t),t) \; ,
\label{eq:sample_ODE}
\end{align}
with a velocity field $v_\theta(x(t),t)$ encoded in a neural
network. This time evolution of the individual events is related to
the time evolution of the underlying density via the continuity
equation
\begin{align}
\frac{\partial p(x,t)}{\partial t} + \nabla_x \left[ p(x,t) v_\theta(x,t) \right] = 0 \; .
\label{eq:continuity}
\end{align} 
The learning task is then to find a velocity field that transforms the
density $p(x,t)$ such that
\begin{align}
 p(x,t) \to 
 \begin{cases}
  \psimp(x) \quad & t \to 0 \\
  \psimr(x)  \quad & t \to 1  \;.
\end{cases} 
\label{eq:didi_limits}
\end{align}
Such a velocity field can be constructed by building on
event-conditional velocity fields. For a given particle-level event $x_0
\sim \psimp(\xp)$, the algorithm samples a corresponding reco-level
event $x_1 \sim \psimr(\xr | \xp=x_0)$, and the two are connected with
a linear trajectory
\begin{align}
    x(t|x_0) 
    = (1-t) x_0 + t x_1
\to \begin{cases}
      x_0 \qquad & t \to 0 \\
      x_1 \sim p(\xr | \xp=x_0)\qquad & t \to 1 \;.
\end{cases}  
\label{eq:didi_trajectory}
\end{align}
Differentiating this trajectory defines the conditional velocity field
\begin{align}
   v(x(t|x_0),t| x_0) 
%   &= \frac{d x(t|x_0)}{dt}\notag \\
   &= \frac{d}{dt} \left[ (1-t) x_0 + t x_1 \right] = - x_0 + x_1 \; .
\label{eq:conditional_velocity}
\end{align}
This is not yet useful as an unfolding network, as it can only unfold to
a pre-specified hard event.  The desired unconditional velocity field
can be obtained via
\begin{align}
    v(x ,t)  = \int dx_0 \; \frac{v(x,t|x_0)p(x,t|x_0)\psimp(x_0)}{p(x,t)} \; ,
    \label{eq:velocity}
\end{align}
where $p(x,t|x_0)$ is the conditional density defined via sampling
from equation~\eqref{eq:didi_trajectory} and $p(x,t)$ is obtained by
integrating out the condition $x_0$. In practice, it is sufficient to
train on fixed data pairs $(\xp, \xr)$ instead of resampling the
posterior $p(\xr | \xp=x_0)$ in each epoch. The velocity field can be
learned from data as a simple regression task with the MSE loss
\begin{align}
    \loss_\text{DiDi} &= \bigl\langle \left[ v_\theta((1-t)x_0+t x_1,t) - (x_1 - x_0)\right]^2 \bigr\rangle_{t \sim \mathcal{U}([0,1]),(x_0, x_1) \sim p(\xp, \xr)} \; .
\label{eq:didi_loss}
\end{align}
Once the network is trained, a reco-level event $x_1 \sim p(\xr)$ can
be transferred by numerically solving the coresponding ODE in
Eq.\eqref{eq:sample_ODE}
\begin{align}
%\frac{d}{dt}x(t) &= v_\theta(x(t),t) 
%\qquad \text{with} \quad x_1 = x(t=1) \notag \\
%\Rightarrow \qquad 
x_0 &= x_1 - \int_0^1 v_\theta (x(t),t) dt
\; .
\label{eq:ODE_solution}
\end{align}

\paragraph{Unpaired DiDi}

The starting premise of most unfolding methods is that the forward model $p(\xr|\xp)$ is known, within uncertainty.  
There may be cases where it is not known~\cite{Howard:2021pos} and instead of pairs $(\xp,\xr)$, we only 
have access to the marginals $\{\xp\}$, $\{\xr\}$.  There is no unique solution to this problem even if the 
detector response is deterministic; however, we can proceed by assuming that the function corresponds to 
the optimal transport map. We consider a variation of DiDi for this configuration by droping the pairing
information between training events~\cite{Butter:2023ira}. This can be
achieved by modifying the conditional trajectory so that $x_1$ is
sampled independently of $x_0$, so Eq.\eqref{eq:didi_trajectory}
becomes
\begin{align}
    x(t|x_0) 
    = (1-t) x_0 + t x_1
\to \begin{cases}
      x_0 \qquad & t \to 0 \\
      x_1 \sim p(\xr)\qquad & t \to 1 \; .
\end{cases}  
\label{eq:didi_trajectory_unpaired}
\end{align}
The loss function is
\begin{align}
    \loss_\text{DiDi-U} &= \bigl\langle \left[ v_\theta((1-t)x_0+t x_1,t) - (x_1 - x_0)\right]^2 \bigr\rangle_{t \sim \mathcal{U}([0,1]),x_0 \sim p(\xp), x_1 \sim p(\xr)} \; .
\label{eq:didi_loss_unpaired}
\end{align}
During training we now sample events independently of each other, and
the learned map will be purely determined by the network and its
training. 

\paragraph{Bayesian network}

Because the distribution mapping loss function does not have a straightforward interpretation as a likelihood, it cannot be simply transformed into
a Bayesian network from first principles. However, we can add the
relevant features of a Bayesian network, as for the
CFM~\cite{Butter:2023fov, Butter:2023ira}. This includes Bayesian
layers, Gaussian distributions of all or some network parameters, and
a KL-term regularizing the network parameters towards a Gaussian
prior,
\begin{align}
    \loss_\text{B-CFM} &= \XLangle \loss_\text{CFM}\XRangle_{\theta \sim q(\theta)} + c  \kl[q(\theta),p(\theta)] \; .
%    &= \XLangle \big( v_t^\theta(x_t) - v_t(x_t;x_0,\epsilon) \big)^2\XRangle_{t\sim U([0,1]),x_0\sim \pd(x_0), \epsilon \sim \normal(0,1)}......\notag \\
\label{eq:BCFM_objective}
\end{align}
The factor $c$ balances the deterministic loss with the Bayesian-inspired
regularization. If the network loss follows from a
likelihood, this factor is fixed by Bayes' theorem. In all other cases
it is a hyperparameter. We have checked that the network
performance as well as the extracted posteriors are stable when
varying $c$ over several orders of magnitudes, suggesting that the 
learned weight distribution corresponds to an inherent property of the setup.

%%%%%%%%%%%%%%%%%%%%%%%%%%%%%%%%%%%%%%%%%%%%%%%%%%%
\subsection{Generative unfolding: cINN, Transfermer, CFM, TraCFM, Latent Diffusion}
\label{sec:ml_gen}

Generative unfolding uses conditional generative networks to learn the
conditional probability describing the inverse simulation
$\pmd(\xp|\xr)$,
\begin{alignat}{9}
  & \psimp
%  \quad \xleftrightarrow{\text{unfolding inference}} \quad 
  && p_\text{unfold}(\xp)
  \notag \\
  & \hspace*{-9mm} \text{\footnotesize paired data} \Bigg\updownarrow
  && \hspace*{+6mm} \Bigg\uparrow {\scriptstyle \pmd(\xp|\xr)}
  \notag \\
  & \psimr 
  \quad \xleftrightarrow{\text{\; correspondence \;}} \quad 
  && \pd(\xr)
\label{eq:schematic_gen}
\end{alignat}
Building a forward surrogate network ($p(\xr|\xp)$) uses the same data and has nearly the same setup as going backwards (($p(\xp|\xr)$)).  The usual assumption of unfolding is that the detector response is universal, which breaks the symmetry of the forward and backwards networks via Bayes' theorem,
\begin{align}
  p(\xp|\xr) = p(\xr|\xp) \; \frac{p(\xp)}{p(\xr)} \; .
\end{align}
For the forward simulation, we assume that the condition on $\xp$ does
not induce a significant prior for the generated $\psimr$. For the
inverse simulation, this prior dependence is relevant and it formally implies that there is no
notion of unfolding single events, even though the generative unfolding
tools provide the corresponding conditional probabilities.

Technically, we start from a simple latent distribution, where the
generative network transforms the required phase space
distribution,
\begin{align}
  z \sim \pl(z)
  \quad 
  \longrightarrow
  \quad
  \xp \sim \pmd(\xp|\xr) \; .
  \label{eq:generative_unfolding}
\end{align}
The phase space distribution of an unfolded dataset is then given as
\begin{align}
\label{eq:convolution}
  \punf(\xp) = \int d \xr \; \pmd(\xp|\xr) \; \pd(\xr) \; .
\end{align}
This approach is based on posterior distributions for individual events, which
means that we can also take single measured events and run them through the model any
number of times.  In practice, Eq.~\ref{eq:convolution} is achieved by sampling from $\pmd(\xp|\xr)$ for data events that follow $\pd(\xr)$ and then ignoring the $\xr$ argument from the resulting dataset of pairs $(\xp,\xr)$.  If we wanted to sample further from the result and/or iterate the procedure, we would need to do something like the second step of OmniFold, which is a local averaging as done for generative models in Ref.~\cite{Backes:2022vmn}.
A key ingredient to unfolding with generative
networks~\cite{Bellagente:2019uyp} is to either train this network with a
likelihood loss~\cite{Bellagente:2020piv}, like for the cINN, or to guarantee the 
probabilistic interpretation through the mathematical setup, like
in the CFM.

%%%%%%%%%%%%%%%%%%%%%%%%%%%%%%%%%%%%%%%%%%%%%%%%%%%
\subsubsection{Conditional INN}
\label{sec:inn_cond}

The original generative network used for unfolding is a normalizing flow~\cite{rezende2016variational}
in its conditional invertible neural network (cINN)
variant~\cite{cinn,Bellagente:2020piv2}. It defines the mapping between the
latent and phase space as an invertible function, conditioned on the
reco-level event,
\begin{align}
z \sim \pl(z)
  \quad 
  \stackrel[\leftarrow \; G^{-1}_\theta(\xp ; \xr)]{G_\theta(z ; \xr) \; \rightarrow}{\xleftrightarrow{\hspace*{1.5cm}}}
  \quad
  \xp \sim \pmd (\xp|\xr) \; .
   \label{inn_idea}
\end{align}
The bijection form allows us to write down the learned density as
\begin{align}
    \pmd (\xp|\xr) = \pl(G^{-1}_\theta(\xp ; \xr)) \left| \text{det}\frac{\partial G^{-1}_\theta(\xp ; \xr)}{\partial \xp} \right|  \; .
   \label{inn_likelihood}
\end{align}
Having access to the network likelihood enables us to use it directly as
loss function and train via maximum likelihood estimation
\begin{align}
    \loss_\text{cINN} &= -\bigl\langle \log \pmd (\xp|\xr) \bigr\rangle_{(x_0, x_1) \sim p(\xp, \xr)} \; .
\label{eq:inn_loss}
\end{align}
This approach requires a bijective map that is flexible enough to
model complex transformations, while still allowing for efficient
computation of the Jacobian determinant. We employ coupling
blocks~\cite{Bellagente:2020piv2}, but replace the affine coupling
blocks with the more flexible rational quadratic spline
blocks~\cite{durkan2019cubicspline}.

\paragraph{Transformer-cINN}

We also consider a transformer extension to the standard cINN~\cite{Heimel:2023mvw}. The architecture 
translates a sequence of reco-level momenta into a sequence of
particle-level momenta. A transformer network encodes the correlations
between all event dimensions at particle level as well as their
correlation with the reco-level event. A small 1D-cINN then generates
the hard-level momenta conditioned on the transformer output. To
guarantee invertibility and a tractable Jacobian, the likelihood and
the generation process are factorized autoregressively
\begin{align}
    \pmd(\xp|\xr) = \prod_{i=1}^n
    \pmd(\xp^{(i)} | c(\xp^{(0)}, \ldots, \xp^{(i-1)}, \xr)) \; .
    \label{eq:transfermer_prob}
\end{align}
The product in Eq.\eqref{eq:transfermer_prob} covers all dimensions at particle level. The function $c$
is learned by the transformer to encode the information about the
reco-level momenta as well as the already generated hard-level
momenta. The one-dimensional conditional densities are encoded in the
cINN. Note that in contrast to Ref.~\cite{Heimel:2023mvw}, this so-called
transfermer is autoregressive in individual one-dimensional
components, instead of in the four momenta grouped by particles.

%%%%%%%%%%%%%%%%%%%%%%%%%%%%%%%%%%%%%%%%%%%%%%%%%%%%%%%%%%%%%%%%%%%%%%%%
\subsubsection{Conditional Flow Matching}
\label{sec:mod_fm}

As an alternative generative network, we employ a diffusion
approach called Conditional Flow Matching~(CFM)~\cite{lipman2023flow,
  Butter:2023fov}. The mathematical structure is the same as for the
DiDi network introduced in Sec.~\ref{sec:diff}. The key difference here is
that the CFM now samples from a Gaussian latent distribution,
conditional on a reco-level event,
Eq.\eqref{eq:generative_unfolding}. This means the time-evolving
density is conditional and interpolates between the boundary
conditions
\begin{align}
 p(x,t|\xr) \to 
 \begin{cases}
  p(\xp|\xr) \qqquad & t \to 0 \\
   \normal(x;0,1) \qqquad & t \to 1  \; ,
\end{cases} 
\label{eq:fm_limits}
\end{align}
while the ODE now reads
\begin{align}
    \frac{dx(t)}{dt} \equiv v_\theta(x(t),t| \xr) \; .
\label{eq:sample_ODE_CFM}
\end{align}
The information about the reco-level event to unfold is no longer
encoded in the initial condition of the ODE, but in an additional
input to the network that predicts the velocity field. Again we start
with paired training data, $x_0 \sim p(\xp)$ and $x_1 \sim p(\xr |
\xp=x_0)$, and define a simple conditional trajectory towards Gaussian
noise,
\begin{align}
    x(t|x_0, \xr) 
%    = \beta_t x_0 + \sigma_t\epsilon 
    = (1-t) x_0 + t \epsilon 
\to \begin{cases}
      x_0 \qqquad & t \to 0 \\
      \epsilon \sim \normal(0,1) \qqquad & t \to 1 \; .
\end{cases} 
\label{eq:trajectory_cfm}
\end{align}
The conditional velocity field is defined via the derivative of the
trajectory
\begin{align}
   v(x(t|x_0, \xr),t| x_0, \xr) 
%   &= \frac{d x(t|x_0)}{dt}\notag \\
   &= \frac{d}{dt} \left[ (1-t) x_0 + t \epsilon \right] = - x_0 + \epsilon \; .
\label{eq:conditional_velocity_cfm}
\end{align}
The rest of the derivation follows analogously to the DiDi
derivation. The loss function is given by the MSE
\begin{align}
    \loss_\text{CFM} &= \bigl\langle \left[ v_\theta((1-t)x_0+t
      \epsilon,t, x_1) - (\epsilon - x_0)\right]^2 \bigr\rangle_{t
      \sim \mathcal{U}([0,1]),(x_0, x_1) \sim p(\xp, \xr), \epsilon
      \sim \mathcal{N}} \; .
\label{eq:cfm_loss}
\end{align}
After training, the CFM can unfold by sampling from the latent noise
distribution and solving the ODE in Eq.\eqref{eq:sample_ODE_CFM}
conditioned on the reco-level event we want to unfold. The crucial
difference to DiDi is that this procedure allows us to unfold the same
reco-level event repeatedly, each time from different noise as
starting point, to sample the posterior distribution $\pmd (\xp|\xr)$.

\paragraph{Transformer-CFM}

%------------------------------------------------------
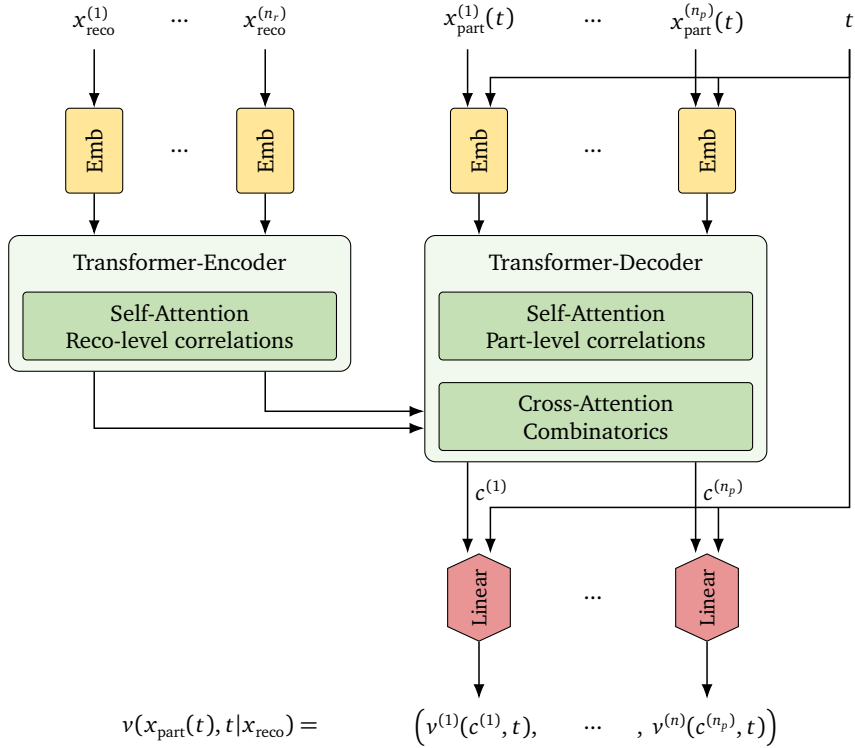
\begin{figure}[t]
    \centering
    \input{tra_cfm}
    \caption{TraCFM architecture, combining the CFM generator with a Transformer 
    encoder-decoder combination to improve combinatorics.}
    \label{fig:tra_cfm}
\end{figure}
%------------------------------------------------------

The velocity field can be encoded in any type of neural network, 
in that sense CFMs do not impose any architectural constraints. While 
linear layers already achieve high precision~\cite{Butter:2023fov, Butter:2023ira}, 
we find that when dealing with complex correlations employing a 
Transformer network further improves results~\cite{Heimel:2023mvw}, 
similar to the INN vs Transfermer case. 

Our TraCFM architecture encoding $v(\xp, t | x_\text{reco})$ 
is shown in Fig.~\ref{fig:tra_cfm}. Its inputs are the reco-level event, 
the intermediate noisy diffusion state $x_\text{part}(t)$ and the 
time $t$. First, each of the reco-level and particle-level dimensions is 
individually mapped into a higher-dimensional embedding space. This 
is done by concatenating the kinematic variable with its one-hot-encoded 
position and filling with zeros up to the specified embedding 
dimension~\cite{Heimel:2023mvw}. For the particle-level dimensions 
we also concatenate the time $t$ to the vector before filling with zeros. 
We experimented with more sophisticated embedding strategies, but found 
no performance improvements. The reco-level embeddings are then fed 
to the transformer encoder, which encodes the correlations among them 
using a self-attention mechanism. The transformer decoder does the same 
for the particle-level dimensions. Finally, the updated embeddings are 
fed to a cross-attention block that learns to resolve the combinatorics 
between reco-level and particle-level objects and outputs a final 
condition $c^{(i)}$ for each particle-level dimension. A single linear layer,
shared between all dimensions, maps this condition together with 
the time $t$ to the individual velocity field components.

To unfold, we start with a sample from the latent distribution as 
$x_\text{part}(t=1)= \epsilon \sim \normal(0,1)$ and solve the ODE 
Eq.~\eqref{eq:sample_ODE_CFM} numerically. Notice that the transformer 
encoder has no time dependence, so we do not need to recalculate it 
at every function call.

\paragraph{Bayesian generative network}

The concept of Bayesian networks can be applied to generative networks
by assigning an uncertainty to the learned underlying phase space
density. This way, the network learns an underlying density to sample
from, and an uncertainty on this density which it can report as an
error of the unit-weight of each generated
event~\cite{Bellagente:2021yyh,Butter:2021csz}. Because the loss of
the normalizing flow is a maximized likelihood, the relation between
the likelihood loss and the regularizing KL-divergence can be derived
from Bayes' theorem.  As an approximation to the full posterior, the 
error bars reported by Bayesian networks are a learned approximation 
to the true uncertainty on the phase space density.

%%%%%%%%%%%%%%%%%%%%%%%%%%%%%%%%%%%%%%%%%%%%%%%%%%%
\subsubsection{Latent Variational Diffusion}
\label{sec:VLD}

To reduce the disparity between different parameterizations of the set of observables 
and to enable a more robust network, Latent Variational 
Diffusion~\cite{Shmakov:2023kjj} introduces a Variational 
Autoencoder to initially map observables from
particle/parton phase space to a
latent space.
This particle encoder learns the mapping 
\begin{align}
  \xp \; \to \;
  z = \textsc{Encoder}_{\text{part}}(\xp ) 
  \in \mathbb{R}^{D_\text{Latent}} \; .
\end{align}
It is implemented as a deep feed-forward network.
This 
latent space can be fine-tuned for the diffusion step, allowing 
enhanced control over the generation process before mapping the result 
back to the observables.

To accommodate variable-length reco-level objects, an additional 
detector encoder maps them to a fixed-length latent vector
\begin{align}
  \xr \; \to \;
  w = \textsc{Encoder}_{\text{reco}}(\xr)
  \in \mathbb{R}^{D_\text{Latent}} \; .
\end{align}
It utilizes a deep feed-forward network for fixed-length inputs 
and a transformer encoder for 
variable-length inputs.

These latent observables provide the inputs for a conditional 
variational diffusion network~\cite{vdm}. VLD employs a continuous-
time, variance-preserving, stochastic diffusion process with a 
noise-prediction parameterization for the score function. The governing stochastic differential equations are
\begin{alignat}{7}
    dz &= f(z_t, t) \, dt + g(t) \, dw 
    &\text{(VLD Forward SDE)} \notag \\
    dz &= \left[ f(z_t, t) - g^2(t) s_\theta(z, w, t) \right] dt + d\bar{w}  
    \qqquad
    &\text{(VLD Reverse SDE)}
\end{alignat}
The drift and diffusion terms are parameterized through a learnable
noise schedule $\gamma_\phi(t)$, which controls the diffusion rate. 
It is encoded in a monotonically increasing deep network as a function of $t$,
\begin{align}
    f(z, t) &= -\frac{1}{2} \left[ \frac{d}{dt} \log \left(1 + e^{\gamma_\phi(t)}\right)\right]z \notag \\
    g(t) &= \sqrt{\frac{d}{dt} \log \left(1 + e^{\gamma_\phi(t)}\right)} \; .
\end{align}
This simplifies the forward process to a time-dependent normal distribution, controlled by $\gamma_\phi(t)$, now interpreted as the
logarithmic signal-to-noise ratio.
\begin{align}
 z_t \sim \mathcal{N} \left(\sigma(-\gamma_\phi(t)) \, z, \sigma(\gamma_\phi(t)) \, \mathbb{I}\right)
 \qquad \text{where} \quad 
 \sigma(x) = \sqrt{\frac{1}{1 + e^{-x}}} \; .
\end{align}
The diffusion score is parameterized via a noise-prediction network,
\begin{align}
 s_\theta(z, w, t) 
 = \frac{\hat{\epsilon}_\theta(z_t, w, t)}{\sigma(-\gamma_\phi(t))} \; ,
 \end{align}
 trained to predict the sampled noise used to generate the forward 
 sample from the diffusion process~\cite{Shmakov:2023kjj, vdm}. 
 It is implemented as a deep feed-forward network, concatenating the 
 three inputs before processing.

Finally, a decoder transforms the initial noisy latent particle
representation back into phase space observables,
\begin{align}
  z_0 
  \; \to \; 
  \hat{x}_\text{part} = \textsc{Decoder}(z_0) \; .
\end{align}
It is again implemented as a deep feed-forward network and outputs 
real-valued estimates of the observables.

All networks are trained in a end-to-end fashion using a unified loss 
which allows the encoders and decoders to fine-tune the latent space 
to the diffusion process, while accurately reconstructing the 
observables. We use a standard normal distribution as the prior over 
the final noisy latent vector, $p(z_1) \sim \mathcal{N}(\textbf{0}, \mathbb{I})$. 
The denoising network, $\gamma_\phi(t)$, is trained to minimize the variance of the following loss term, while all other networks are trained to minimize its expectation value:
\begin{alignat}{7}
  \loss_\text{VLD} 
  &= \kl [q(z_1 | x),p(z_1)] 
  &\text{(Prior Loss)} \notag \\ 
  &+ \XLangle \norm{\textsc{Decoder}(z_0) - x}_2^2 \XRangle_{q(z_0 | x)} 
  &\text{(Reconstruction Loss)} \notag \\
 &+ \XLangle  \gamma_\phi'(t) \norm{\epsilon - \hat{\epsilon}_\theta(z, w, t)}_2^2
 \XRangle_{\epsilon \sim \mathcal{N}(\textbf{0}, \mathbb{I}), t \sim \mathcal{U}(0, 1)} 
 \qqquad &\text{(Denoising Loss)} 
 %\\
 % &+ \mathbb{E}_{q(z_0 | x)} \left [ \norm{\textsc{Decoder}(z_0) - x} \right ]_2^2 
 % &\text{(Reconstruction Loss)} \notag \\
 %&+ \mathbb{E}_{\epsilon \sim \mathcal{N}(\textbf{0}, \mathbb{I}), t \sim \mathcal{U}(0, 1)} \left [ \gamma_\phi'(t) \norm{\epsilon - \hat{\epsilon}_\theta(z, w, t)}_2^2\right ] 
 %\qqquad &\text{(Denoising Loss)}
\end{alignat}

\clearpage
%%%%%%%%%%%%%%%%%%%%%%%%%%%%%%%%%%%%%%%%%%%%%%%%%%%
\section{Detector unfolding: $Z$+jets}
\label{sec:omni}

%%%%%%%%%%%%%%%%%%%%%%%%%%%%%%%%%%%%%%%%%%%%%%%%%%%
\subsection{Data and preprocessing}
\label{sec:omni_pre}

As a first test case for the various ML-Unfolding methods we use a new, bigger version of
the public dataset from Ref.~\cite{Andreassen:2019cjw}, now available at Ref.~\cite{benjamin_2024_10668638}\footnote{For a comparison with 
classical unfolding methods, we refer to Refs.~\cite{Andreassen:2019cjw} and~\cite{Backes:2023ixi}.}. The events describe 
\begin{align} 
 pp \to Z + \text{jets}
\end{align}
production at $\sqrt{s} = 14$~TeV, simulated with Pythia~8.244~\cite{Sjostrand:2014zea} 
with Tune~26. 
In contrast to the original dataset, detector
effects are now simulated with the updated Delphes~3.5.0~\cite{deFavereau:2013fsa},
and the CMS tune, that uses particle flow
reconstruction. The jets are clustered using all particle flow
objects available at detector level and all stable non-neutrino truth
particles at particle level. Jets are defined by the
anti-$k_T$ algorithm~\cite{Cacciari:2008gp} with 
$R= 0.4$, as implemented in FastJet~3.3.2~\cite{Cacciari:2011ma}. 
The dataset contains around 24M simulated events, 20M for training and 4M for testing.

%------------------------------------------------------
\begin{figure}[b!]
    \includegraphics[width=0.33\textwidth, page=1]{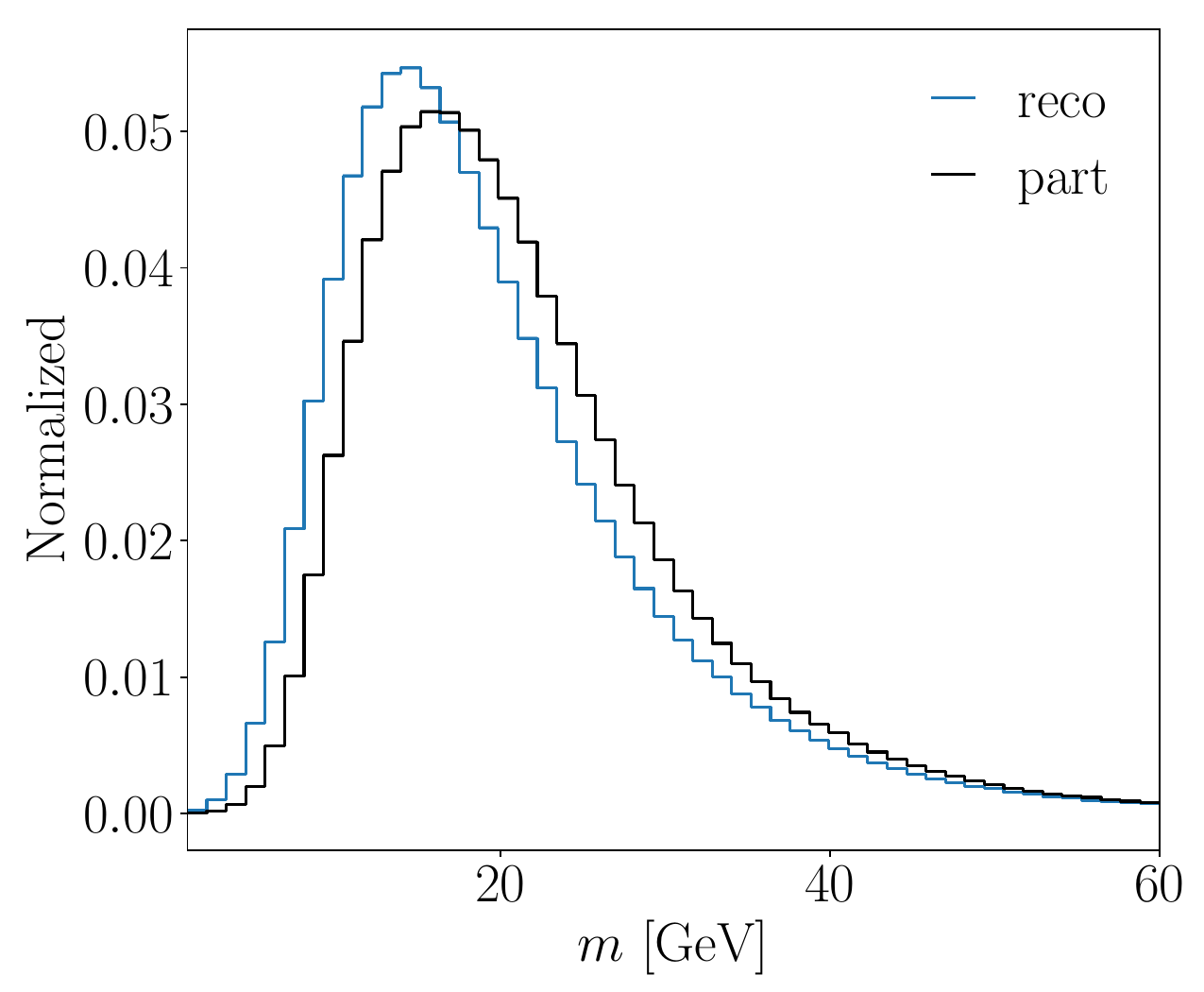}
    \includegraphics[width=0.33\textwidth, page=2]{figs/new_dataset/new_dataset.pdf}
    \includegraphics[width=0.33\textwidth, page=3]{figs/new_dataset/new_dataset.pdf}\\
    \includegraphics[width=0.33\textwidth, page=4]{figs/new_dataset/new_dataset.pdf}
    \includegraphics[width=0.33\textwidth, page=5]{figs/new_dataset/new_dataset.pdf}
    \includegraphics[width=0.33\textwidth, page=6]{figs/new_dataset/new_dataset.pdf}
    \caption{Subjet distributions for the $Z+$jets dataset, at the particle level, and
    at the reco level.}
    \label{fig:dataset2}
\end{figure}
%------------------------------------------------------

We focus on six observables 
describing the 
leading jet: mass $m$, width $\tau_{1}^{(\beta=1)}$, 
multiplicity $N$, soft-drop~\cite{Larkoski:2014wba,Dasgupta:2013ihk} mass
$\rho = m_\text{SD}^{2} / p_T^2$ and momentum
fraction $z_g$ using $z_\text{cut}= 0.1$ and $\beta = 0$, and 
the $N$-subjettiness ratio $\tau_{21} = \tau_{2}^{(\beta=1)} /
\tau_{1}^{(\beta=1)}$~\cite{Thaler:2010tr}. 
For 0.8\% of the events we map an undefined jet groomed mass
$\log\rho$ or N-subjettiness ratio $\tau_{21}$ to zero.

The distributions are shown in Fig.~\ref{fig:dataset2}. We apply a dedicated 
preprocessing to the jet multiplicity and the groomed momentum fraction. 
The jet multiplicity is an integer feature, which forces the network to interpolate, 
so we smooth the distribution by adding uniform noise $u\sim\mathcal{U}[-0.5,0.5)$.
This preprocessing can be inverted. The groomed momentum fraction 
features a discrete peak at $z_g=0$ and sharp cuts at $z_g=0.1$ and $0.5$.
We move the peak to $z_g=0.097$ and add uniform noise $u\sim\mathcal{U}[0,0.003)$. 
Next, we take the logarithm to make the distribution more uniform.  
We then shift and scale the distribution to stretch from $-1$ to $+1$ and take 
the inverse error function to transform its shape to an approximate normal distribution. 
Finally, all six observables are standardized by subtracting the means and 
dividing by the standard deviations.

Also in Fig.~\ref{fig:dataset2}, we show the effect of the detector simulation. They
are most significant for the jet multiplicity, the groomed jet mass, and the $N$-subjettiness
ratio. All these shifts are driven by the finite energy threshold of the detector.

%%%%%%%%%%%%%%%%%%%%%%%%%%%%%%%%%%%%%%%%%%%%%%%%%%%
\subsection{Reweighting}
\label{sec:omni_omni}

As in Sec.~\ref{sec:ml}, we start
with the OmniFold reweighting on the $Z+\text{jets}$
dataset. We then introduce the Bayesian version (bOmniFold) and
compare their performance. We train both networks for two
different unfolding tasks. First, we evaluate their performance on
the same dataset as the other generative networks, splitting it in two
halves, but adding noise to one of them as described below. 
Then, we go back to the previous Pythia
dataset
and task the classifiers with learning
the likelihood ratio between Pythia and Herwig. We use this ratio to
reweight Herwig onto Pythia. 

%%%%%%%%%%%%%%%%%%%%%%%%%%%%%%%%%%%%%%%%%%%%%%%%%%%
\subsubsection{Training on Pythia with added noise}

%------------------------------------------------------
\begin{figure}[b!]
    \includegraphics[width=0.33\textwidth, page=1]{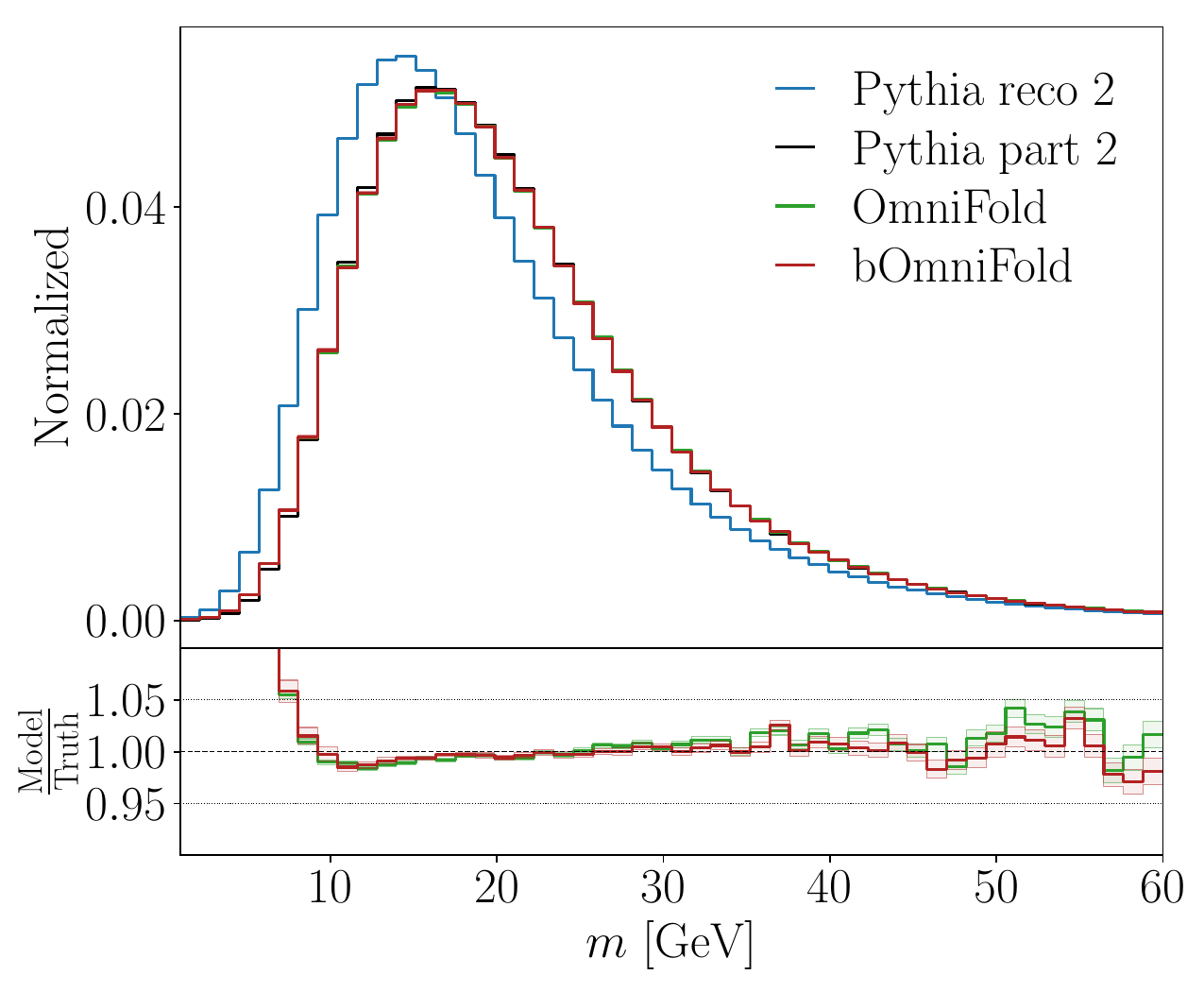}
    \includegraphics[width=0.33\textwidth, page=2]{figs/Omnifold/PtoP/PtoP-observables.pdf}
    \includegraphics[width=0.33\textwidth, page=3]{figs/Omnifold/PtoP/PtoP-observables.pdf} \\
    \includegraphics[width=0.33\textwidth, page=4]{figs/Omnifold/PtoP/PtoP-observables.pdf}
    \includegraphics[width=0.33\textwidth, page=5]{figs/Omnifold/PtoP/PtoP-observables.pdf}
    \includegraphics[width=0.33\textwidth, page=6]{figs/Omnifold/PtoP/PtoP-observables.pdf}
    \caption{Unfolded distributions from event reweighting using OmniFold and
      bOmniFold. The bOmniFold error bar is based on 
      drawing 20 Bayesian samples. For OmniFold the error bar represents the bin-wise statistical uncertainty.}
    \label{fig:omnifold-PtoP-observables}
\end{figure}
%------------------------------------------------------

For this section, we employ the combination of Pythia with the updated
Delphes version. We merge the training and test sets with 24.3M
events, of which we use 10.9M for training, 1.2M for validation, and 12.2M for testing. In each of these splits, we label half of the events as Pythia~1 and the other half as Pythia~2. 
The classifier has to learn to reweight Pythia~1 onto Pythia~2. 

If we train (b)Omnifold on this task, it will just learn a constant 
classifier value of 
0.5, so we add 
Gaussian noise
$\varepsilon\sim\mathcal{N}(0,1)$ to each of the raw features before
preprocessing, scaled by the standard deviation of the respective
feature $\sigma_{x}$ and an additional custom factor $f$ to modify the
relative importance of the noise,
\begin{align}
    x \rightarrow \tilde{x} = x + f \cdot \sigma_{x} \varepsilon
    \qquad \text{with} \qquad 
    \varepsilon\sim \mathcal{N}(0,1) \; ,
\end{align}
where we use $f=0.1$.

We train OmniFold
(13k parameters) and its Bayesian-network counter part  bOmniFold 
(2$\times$13k parameters) with
identical settings for 30~epochs. 
The unfolded distributions are
shown in Fig.~\ref{fig:omnifold-PtoP-observables}. While this reweighting task might not be realistic, it defines 
an illustrative benchmark for the performance of an unfolding network.
For each of the one-dimensional kinematic distributions, the agreement
between the unfolded and true particle-level events is at the per-cent
level or better.
The only exceptions are sparsely populated tails with
too little training data, or sharp features with limited resolution. 
The differences between the OmniFold and 
bOmniFold results are even smaller. 
A selection of summary statistics are presented in Tab.~\ref{tab:metrics_zjets-PtoP-Omnifold}, where we show the Wasserstein~1-distance, the triangular distance, and the energy distance for the six kinematic observables. The two methods were not separately optimized, we just started with a generic Omnifold setup and supplemented it with the Bayesian network features. Uncertainties on the statistics are not included in these illustrative metrics.

For the uncertainties, 
we see that it tends to cover the deviation of the unfolded 
distributions from the truth target towards 
increasingly sparse tails. Far in the tails, where there is too little 
training data altogether, the networks learn neither the density nor
an error bar on it.  

%------------------------------------------------------
\begin{table}[t]
\resizebox{\textwidth}{!}{
 \begin{tabular}{l|c|c|c}
    \toprule
      & $m\textrm{ }[\mathrm{GeV}]$       & $w$                & $N$        \\ 
      \midrule
      OmniFold   & 0.59098 / \textbf{0.12493} / 13.72203 & 0.01001 / \textbf{1.62601} / 2.99618 & 0.67919 / 0.03034 / 18.47942 \\
      bOmniFold  & \textbf{0.37180} / 0.14208 / \textbf{9.89718} & \textbf{0.00542} / 1.64286 / \textbf{2.24587} & \textbf{0.22693} / \textbf{0.02176} / \textbf{4.97982} \\
      \midrule
      & $\log{\rho}$           & $z_{g}$                     & $\tau_{21}$ \\
      \midrule
      OmniFold & 0.40320 / 0.72494 / 15.60005 & 0.01550 / 15.30356 / 4.81947 & \textbf{0.00931} / 0.02746 / \textbf{1.40143} \\
      bOmniFold  & \textbf{0.12501} / \textbf{0.67605} / \textbf{5.59003} & \textbf{0.01109} / \textbf{15.27470} / \textbf{4.51572} & 0.00956 / \textbf{0.02183} / 1.54405 \\
     \bottomrule
\end{tabular}}
\caption{Metrics evaluating the performance of the different unfolding networks, for each of the 
one-dimensional kinematic distributions. We show the Wasserstein 1-distance ($\times 10$), the triangular distance ($\times 1000$),
and the energy distance $(\times 1000)$.}
\label{tab:metrics_zjets-PtoP-Omnifold}
\end{table}
%------------------------------------------------------

%------------------------------------------------------
\begin{figure}[b!]
    \centering
    \includegraphics[width=0.65\textwidth, page=1]{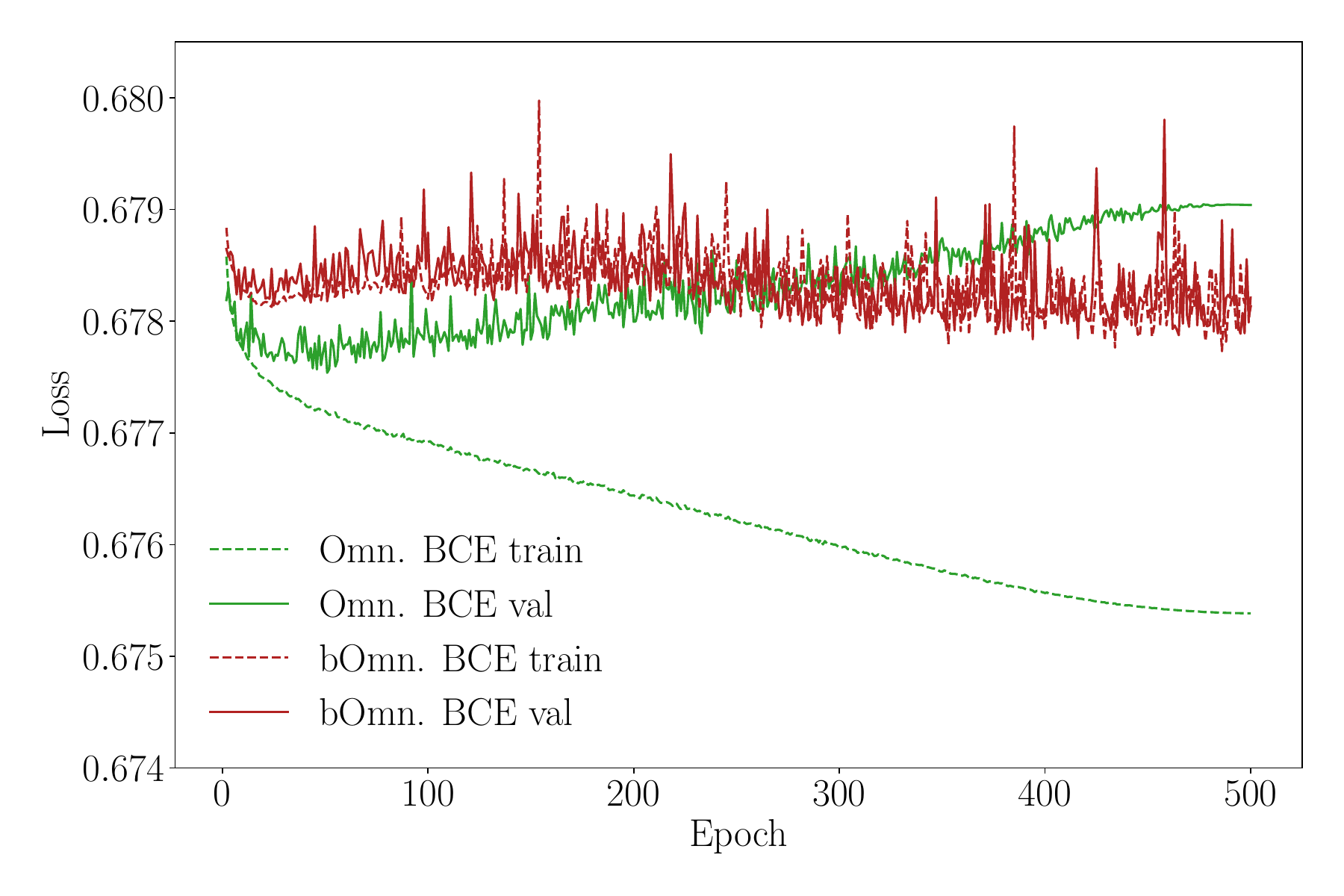}
    \caption{BCE losses during training for 500 epochs for Omnifold (green)
     and bOmnifold (red), for Herwig-to-Pythia reweighting.}
    \label{fig:omnifold-HtoP-losses}
\end{figure}
%------------------------------------------------------

%%%%%%%%%%%%%%%%%%%%%%%%%%%%%%%%%%%%%%%%%%%%%%%%%%%
\subsubsection{Reweighting Herwig onto Pythia}

For a more realistic (b)OmniFold task, we go back to the 
original Pythia dataset, for which we also have a Herwig~\cite{Bewick:2023tfi} version 
with the same version of Delphes, 
as introduced in Ref.~\cite{Andreassen:2019cjw}.
We train 
(b)OmniFold for 500~epochs on 2M events, and test on
664k events~\cite{Diefenbacher:2023wec}.

%------------------------------------------------------
\begin{figure}[t]
    \centering
    \includegraphics[width=0.42\textwidth, page=1]{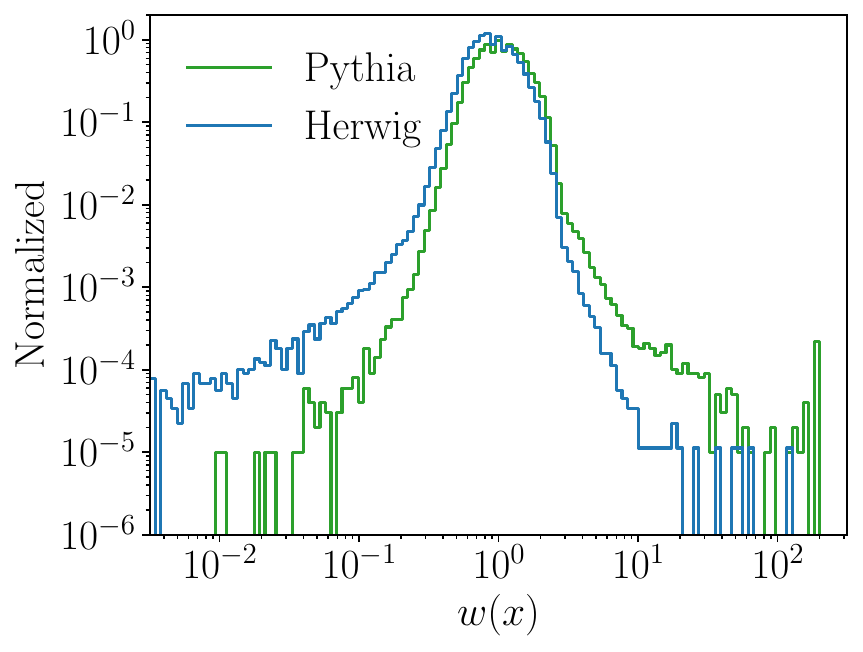}
    \hspace*{0.1\textwidth}
    \includegraphics[width=0.42\textwidth, page=1]{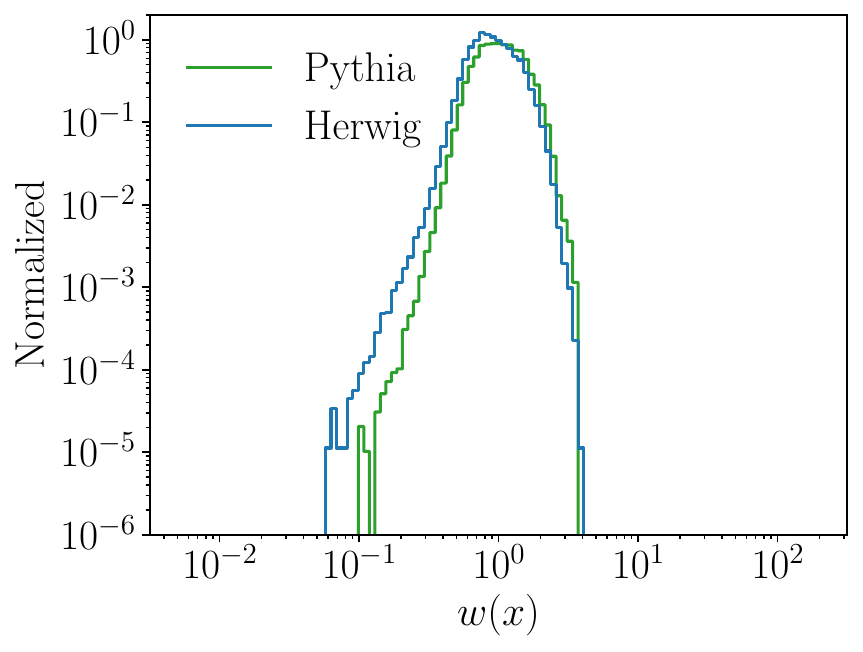}
    \caption{Weight distribution (clipped at 200) in the training set for 
      Herwig-to-Pythia reweighting: OmniFold (left) vs bOmniFold (right). 
      For each network we histogram the weights for the Herwig and Pythia 
      data points.}
    \label{fig:omnifold-HtoP-weights_train}
\end{figure}
%------------------------------------------------------

First, we show the losses as a function of the training in Fig~\ref{fig:omnifold-HtoP-losses}. 
This comparison shows the challenge of the classifier training, which rapidly overtrains after about 20 epochs. This behavior does not appear in the previous study with noisy Pythia events and is due to the 
smaller training data for the Herwig reweighting. For large numbers of
training epochs, the loss on the validation dataset indicates 
a decreasing performance due to overtraining. For applications which require an LHC-level of precision such overtraining may become 
a problem. It can be avoided, for instance, using  regularization techniques, such as
dropout. Both of these mechanisms are part of the Bayesian network
architecture, in case of the regularization with a strength given by 
Bayes' theorem. In Fig.~\ref{fig:omnifold-HtoP-losses}
we see that the bOmniFold training continues to improve even after a large number of epochs, with no overtraining.  
Interestingly, bOmniFold has larger epoch-to-epoch fluctuations and has a worse minimum validation loss than OmniFold,
but does not show signs of overtraining. 
This illustrates a potential tradeoff between accuracy and stability.

As before, the unfolded observables agree well 
between OmniFold and bOmniFold. Because 
of the difference 
between the training data, or prior, and the data we then unfold, 
the true particle-level distributions are not exactly reproduced, 
and this model uncertainty is not intended to be covered by the Bayesian 
error estimate. 

An interesting feature of the bOmniFold training is that it has suppressed tails in the weight distribution with respect to OmniFold, as shown in Fig.~\ref{fig:omnifold-HtoP-weights_train}, even though both networks learn the same reweighting map.  Large and small weights lead to undesired statistical dilution of the dataset and it will be interesting to explore in the future the interplay between statistical dilution and accuracy.

%%%%%%%%%%%%%%%%%%%%%%%%%%%%%%%%%%%%%%%%%%%%%%%%%%%
\subsection{Mapping distributions}
\label{sec:omni_map}

The same subjet unfolding can be tackled with distribution mapping, using  the 
Schr\"odinger Bridge and Direct Diffusion, both introduced in Sec.~\ref{sec:SB}. 
The implementation of the Schr\"odin\-ger Bridge follows the original 
Pytorch~\cite{pytorch}  implementation~\cite{Diefenbacher:2023wec}. The noise 
prediction network is implemented  using a fully connected architecture with 
additional skip connections, specifically using 
six \textsc{ResNet}~\cite{he2016deep} blocks, with each residual layer connected
to the  output of a single MLP through a skip connection. The Bayesian version 
replaces the  original MLPs. The training uses the Adam~\cite{adam} optimizer. The 
total number of trainable parameters is around 2M split equally between the mean 
and and standard deviation of the trainable weights.

%------------------------------------------------------
\begin{figure}[t]
\includegraphics[width=0.33\textwidth, page=1]{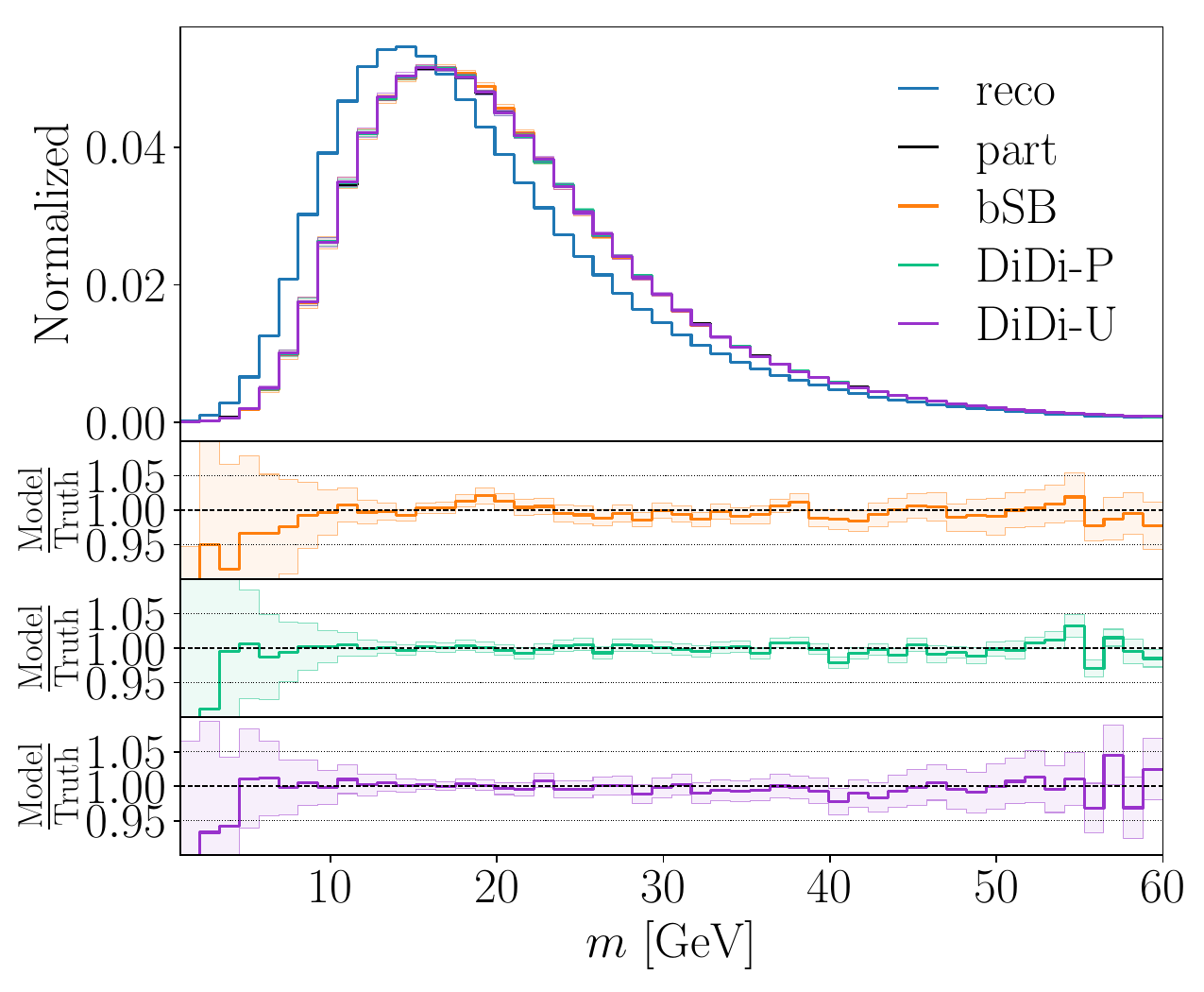}
\includegraphics[width=0.33\textwidth, page=2]{figs/new_dataset/pythia_large/LARGE_maxruntime-observables-SB_DiDi-bay.pdf}
\includegraphics[width=0.33\textwidth, page=3]{figs/new_dataset/pythia_large/LARGE_maxruntime-observables-SB_DiDi-bay.pdf} \\
\includegraphics[width=0.33\textwidth, page=4]{figs/new_dataset/pythia_large/LARGE_maxruntime-observables-SB_DiDi-bay.pdf}
\includegraphics[width=0.33\textwidth, page=5]{figs/new_dataset/pythia_large/LARGE_maxruntime-observables-SB_DiDi-bay.pdf}
\includegraphics[width=0.33\textwidth, page=6]{figs/new_dataset/pythia_large/LARGE_maxruntime-observables-SB_DiDi-bay.pdf}
\caption{Unfolded distributions from distribution mapping, using the Schr\"odinger Bridge 
and DiDi. The Bayesian error bars are based on drawing 50 samples.}
\label{fig:large_maxruntime-observables-SB_DiDi-bay}
\end{figure}
%------------------------------------------------------

During data generation, we sample using the the MAP prediction, \ie fix every network weight at the learned mean. Uncertainties are derived by sampling 50 times from the learned weight distributions. In Fig.~\ref{fig:large_maxruntime-observables-SB_DiDi-bay}, we quantify the agreement between the unfolded and truth one-dimensional kinematic distributions.
The unfolding performance can be compared to the noisy reweighting benchmark in 
Fig.~\ref{fig:omnifold-PtoP-observables}. 
The agreement between unfolded and truth-level observables is still precise to the per-cent level.
Notably, the largest deviations from the truth distribution occur in the low-statistics edges, while the bulks of the distributions are well described by the generative mapping, and the deviations from the truth are well covered by the Bayesian uncertainty.  

An alternative method for the same tasks is Direct Diffusion. We encode the velocity 
field in a standard Bayesian MLP, after not seeing better results with more advanced
networks. Again, we implement the network in Pytorch, train it using the Adam 
optimizer, use the MAP prediction,
and draw 50 Bayesian weight samples to estimate the uncertainties. 
We use the same setup for paired and unpaired DiDi. The only difference is the reshuffling 
of the reco-particle pairings at the beginning of each epoch in the unpaired setting. The 
network size comes to about 3M parameters, 1.5M weights each associated with the mean 
and the standard deviation. 

The results are compared to the Schr\"odinger Bridge in 
Fig.~\ref{fig:large_maxruntime-observables-SB_DiDi-bay}.
Both variants of Direct Diffusion learn the observables with per-cent precision over 
the entire phase space, and better than that in the well-sampled bulk. For the 
central prediction, the paired training data makes the unfolding slightly more
precise and more stable. 

A difference between paired and unpaired DiDi is that the latter 
might be slightly less stable and learns significantly larger Bayesian 
uncertainties. This is consistent over several 
trainings. At the level of kinematic distribution we do not observe any shortcoming
for unfolding through distribution matching, and the difference between 
paired and unpaired training data is minor. We will come back to the conceptual difference in Sec.~\ref{sec:omni_mig}.

%%%%%%%%%%%%%%%%%%%%%%%%%%%%%%%%%%%%%%%%%%%%%%%%%%%
\subsection{Generative unfolding}
\label{sec:omni_gen}

%------------------------------------------------------
\begin{figure}[t!]
\includegraphics[width=0.33\textwidth, page=1]{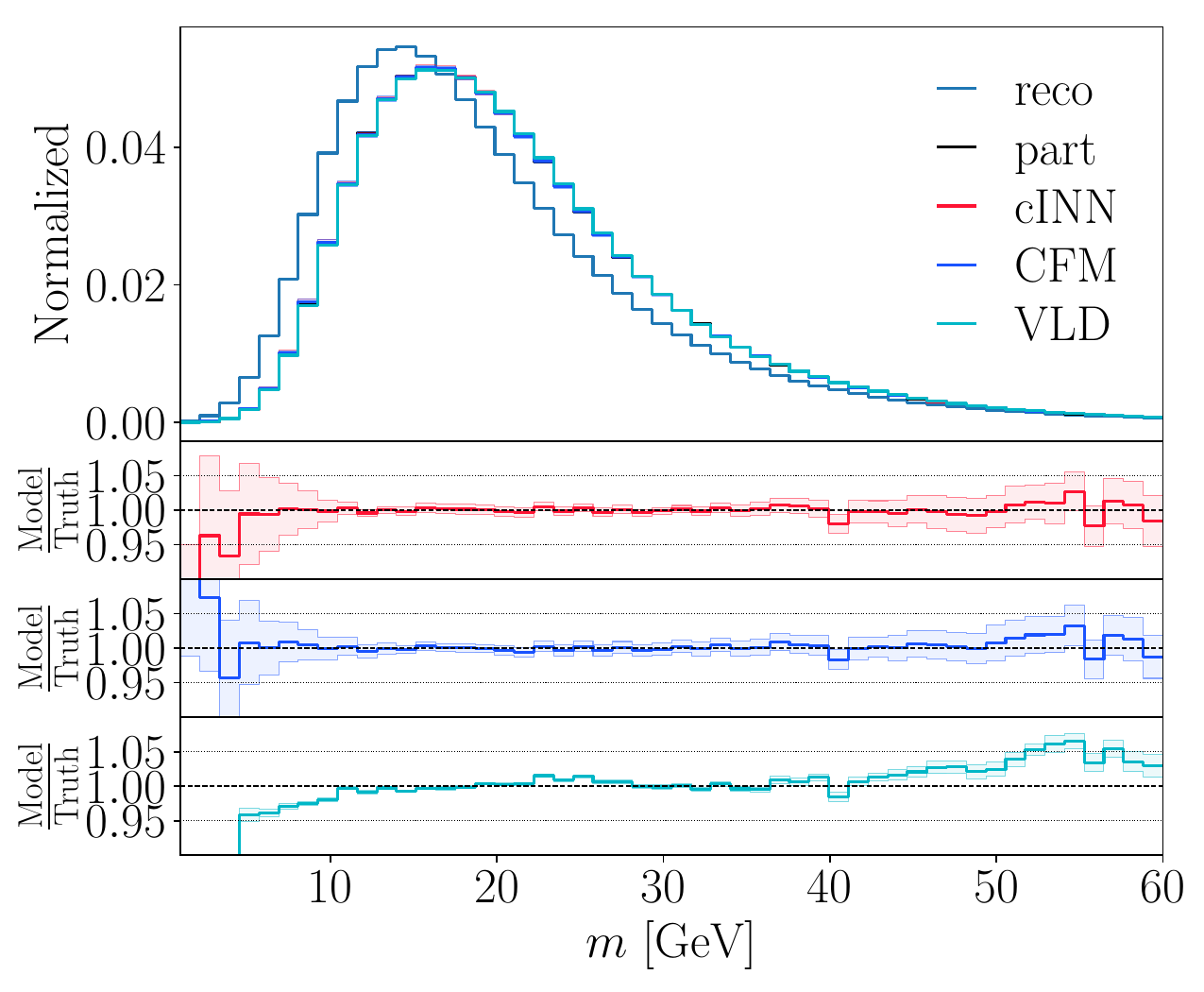}
\includegraphics[width=0.33\textwidth, page=2]{figs/new_dataset/pythia_large/LARGE_maxruntime-observables-CFM_INN_VLD-bay.pdf}
\includegraphics[width=0.33\textwidth, page=3]{figs/new_dataset/pythia_large/LARGE_maxruntime-observables-CFM_INN_VLD-bay.pdf} \\
\includegraphics[width=0.33\textwidth, page=4]{figs/new_dataset/pythia_large/LARGE_maxruntime-observables-CFM_INN_VLD-bay.pdf}
\includegraphics[width=0.33\textwidth, page=5]{figs/new_dataset/pythia_large/LARGE_maxruntime-observables-CFM_INN_VLD-bay.pdf}
\includegraphics[width=0.33\textwidth, page=6]{figs/new_dataset/pythia_large/LARGE_maxruntime-observables-CFM_INN_VLD-bay.pdf}
\caption{Unfolded distributions from conditional generation, using cINN, CFM and VLD. 
For cINN and CFM, the Bayesian errors are based on drawing 50 samples, and the 
MAP estimate is obtained by unfolding each event 30 times. For VLD, we show
the bin-wise mean and standard deviation of 33 unfoldings.} 
\label{fig:large_maxruntime-observables-CFM_INN-bay}
\end{figure}
%------------------------------------------------------

The third unfolding method we study is based on learned conditional probabilities,
as defined in the statistical description of unfolding.
It relies on 
paired training data. Differences appear when we vary the generative network 
architecture used. We skip the original GAN implementation~\cite{Bellagente:2019uyp},
because more modern generative networks have been shown to learn phase space
distributions more precisely~\cite{Butter:2021csz,Butter:2023fov}.
The cINN~\cite{Bellagente:2020piv} is implemented in PyTorch~\cite{pytorch} and uses
the FrEIA library~\cite{freia} with RQS coupling blocks. By default, we use 
its Bayesian version~\cite{Bellagente:2021yyh}, which tracks the uncertainty 
in the learned phase space density as variations of unit event weights.
Our CFM~\cite{Butter:2023fov} encodes the velocity field in the same linear layer 
architecture as DiDi, including the Bayesian version. Predictions are obtained by unfolding each event 30 times with the MAP weights, and the uncertainties are obtained from drawing 50 sets of weights from the Bayesian network.

VLD is implemented and trained using the same JAX 
codebase released alongside Ref.~\cite{Shmakov:2023kjj}. 
Observables are first pre-processed so that each 
marginal distribution follows a standard normal via a 
quantile transform. Predictions are generated using 
the DPM++ multi-step solver~\cite{lu2023dpmsolver} 
with 1000 inference steps and the learned schedule. 
Uncertainty is estimated by sampling each unfolding 33 
times using a different seed for generating the prior 
noise.

In Fig.~\ref{fig:large_maxruntime-observables-CFM_INN-bay} we show the 
results from the cINN, CFM, and VLD. As for the (b)OmniFold reweighting and the 
distribution matching, all kinematic distributions are reproduced at the per-cent
level or better. While the performance of the cINN and the CFM are very similar,
the VLD approach shows slightly larger deviations from the target distributions.

%%%%%%%%%%%%%%%%%%%%%%%%%%%%%%%%%%%%%%%%%%%%%%%%%%%
\subsection{Learned event migration}
\label{sec:omni_mig}

\begin{figure}[t!]
    \includegraphics[width=0.33\textwidth, page=1]{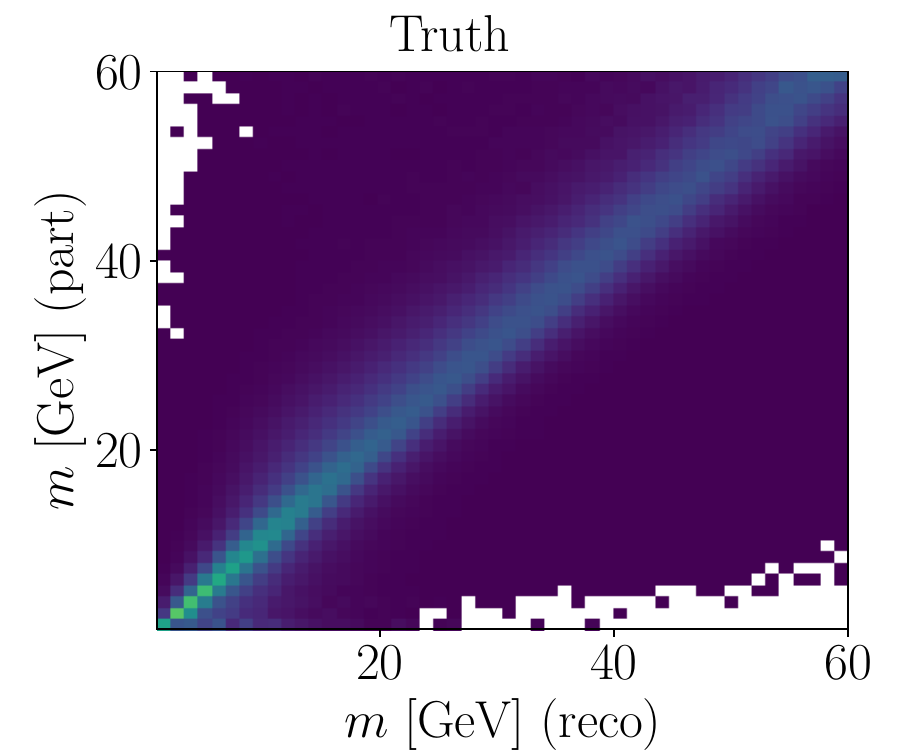}
    \includegraphics[width=0.33\textwidth, page=3]{figs/new_dataset/migration/LARGEv2-gt_migration2.pdf}
    \includegraphics[width=0.33\textwidth, page=6]{figs/new_dataset/migration/LARGEv2-gt_migration2.pdf} \\
    \includegraphics[width=0.33\textwidth, page=1]{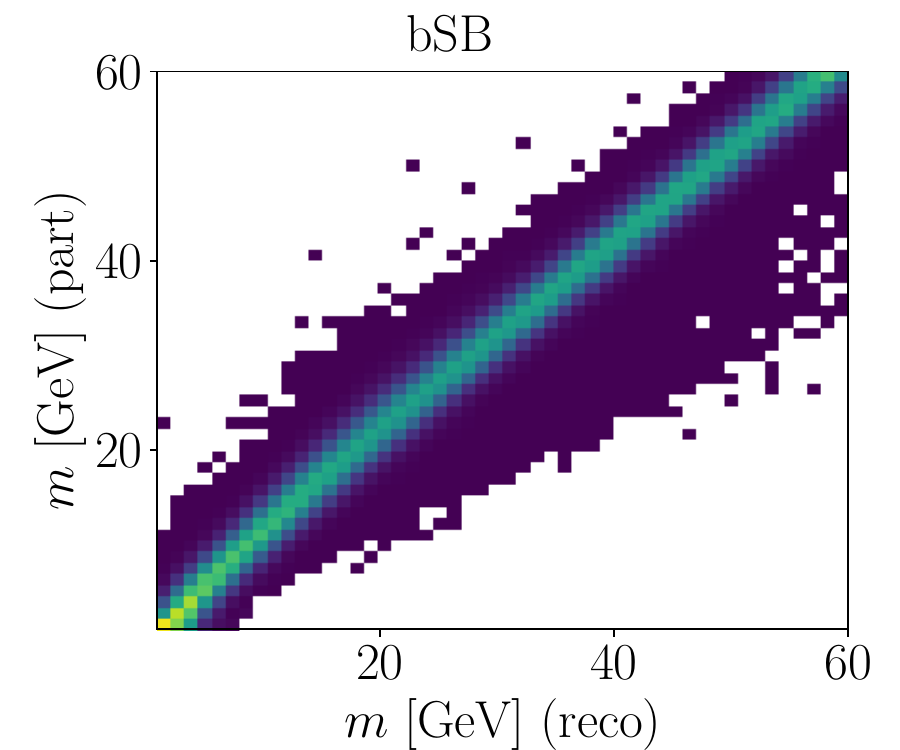}
    \includegraphics[width=0.33\textwidth, page=3]{figs/new_dataset/migration/LARGEv2-SB_migration2.pdf}
    \includegraphics[width=0.33\textwidth, page=6]{figs/new_dataset/migration/LARGEv2-SB_migration2.pdf} \\
    \includegraphics[width=0.33\textwidth, page=1]{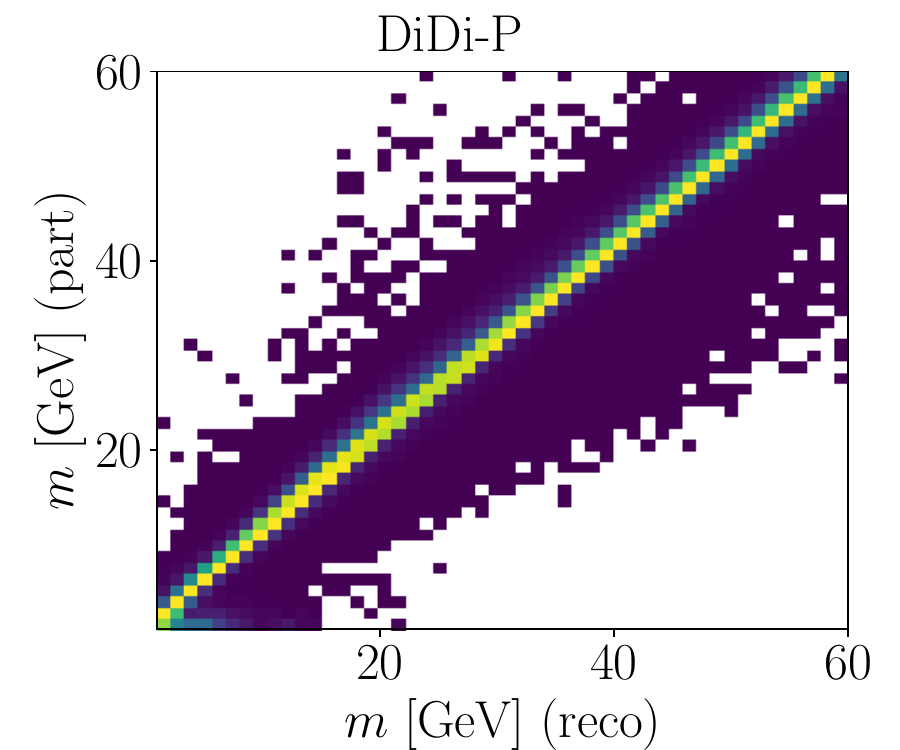}
    \includegraphics[width=0.33\textwidth, page=3]{figs/new_dataset/migration/LARGEv2-DiDi_P_migration2.pdf}
    \includegraphics[width=0.33\textwidth, page=6]{figs/new_dataset/migration/LARGEv2-DiDi_P_migration2.pdf} \\
    \includegraphics[width=0.33\textwidth, page=1]{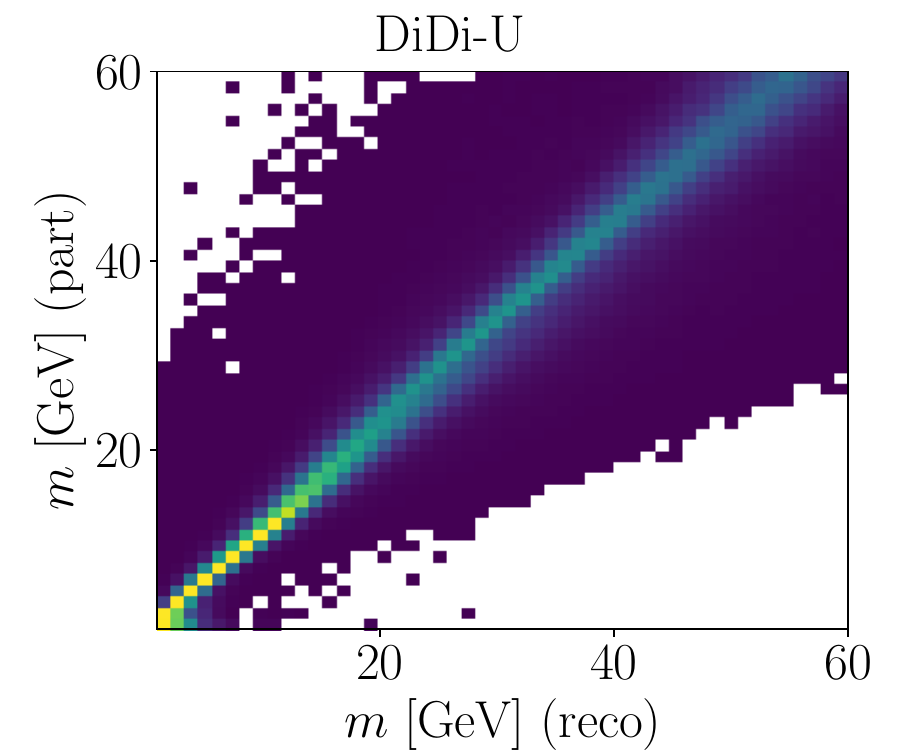}
    \includegraphics[width=0.33\textwidth, page=3]{figs/new_dataset/migration/LARGEv2-DiDi_U_migration2.pdf}
    \includegraphics[width=0.33\textwidth, page=6]{figs/new_dataset/migration/LARGEv2-DiDi_U_migration2.pdf} \\
    \includegraphics[width=0.33\textwidth, page=1]{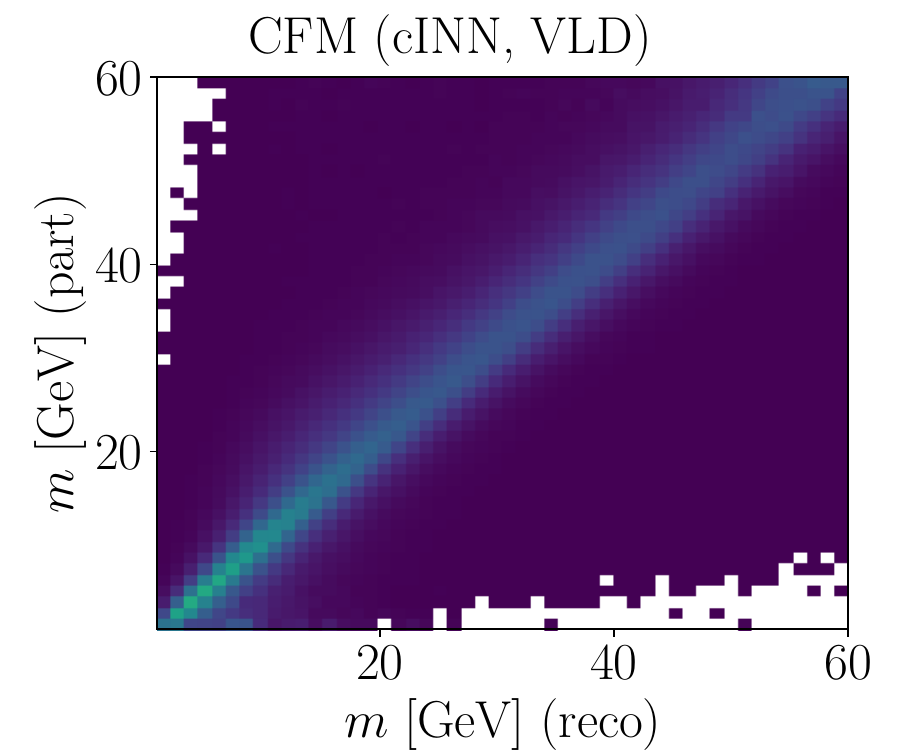}
    \includegraphics[width=0.33\textwidth, page=3]{figs/new_dataset/migration/LARGEv2-CFM_migration2.pdf}
    \includegraphics[width=0.33\textwidth, page=6]{figs/new_dataset/migration/LARGEv2-CFM_migration2.pdf}
    \caption{Migration maps for 
      three representative distributions.
      From the top: forward detector simulation, Schr\"odinger
      Bridge, paired DiDi, unpaired DiDi, and CFM/cINN/VLD, which
      are all looking identical. The bin 
      contents are normalized such that
      each row sums to one.}
    \label{fig:large-optimal_transport}
\end{figure}

For the generative network methods, it can be instructive to examine the learned map between reco events and truth events. In the top panels of Fig.~\ref{fig:large-optimal_transport}
we start with the migration described by the paired events from the 
forward detector simulation. We show three of the kinematic distributions
defined in Fig.~\ref{fig:dataset2}. The results for the other distributions
lead to the same conclusions. For the jet mass, the multiplicity, and $\tau_{21}$
we see that the optimal transport defined by the detector is quite noisy. 
While for the jet mass most events form a diagonal with little bias, 
the jet multiplicity shows a linear correlation with a non-trivial slope, 
and $\tau_{21}$ indicates a saturation effect in the forward direction $\xp \to \xr$,
such that $\tau_{21}(\xr)$ does not reach one.

In the next 
row we show the results from the Schr\"odinger Bridge, which 
is similar, but slightly noisier than DiDi trained on paired events, shown in the third row. 
Using paired events, these generative networks 
learn a very efficient diagonal transport map, with a spread that 
is more narrow than the actual detector. The main features of the 
detector truth are reproduced well. 

Next, we see that the unpaired DiDi network again learns an efficient transport map, 
but with a significantly broader spread than the same network trained on paired 
events. The reason is that 
ignoring the 
event pairing leads to a noisier training, but again
reproducing the main features of the detector. We emphasize that 
unpaired training seems to bring DiDi-like implementations closer to 
describing the actual detector, but this is an artifact in that the detector 
mapping is noisier than the optimal transports from distribution mapping, 
and training on unpaired samples is also noisier, but the two are not positively
related.

Finally, we show the transport learned by the 
conditional CFM networks. Not shown are the corresponding cINN and VLD results, which are visually identical to the CFM results. 
The conditional generative models indeed learn 
the correct detector transport from the
paired events, indicating that conditional generative networks
indeed encode the conditional probabilities from Eq.\eqref{eq:schematic_gen}. \\
Finally, there is the question if the learned distributions have failure modes that cannot be seen from the marginal distributions. To answer it, 
we systematically search
for mis-matched correlations using a trained classifier between the true training 
events and the same number of unfolded events. 
The
different methods give ROC-AUC values in the range of 0.500-0.55. Especially for the 
cINN and CFM networks the AUC is consistent with 1/2, with a tiny number of 
statistical outliers. This indicates that despite the different transport maps shown in Fig.~\ref{fig:large-optimal_transport}, all of the networks faithfully reproduce the target distribution with percent level precision.
Remaining differences  exist, but they are consistent with noise and cannot be traced back to a mis-modeled feature in the underlying phase space density.

%------------------------------------------------------
%------------------------------------------------------

\clearpage
%%%%%%%%%%%%%%%%%%%%%%%%%%%%%%%%%%%%%%%%%%%%%%%%%%
\section{Unfolding to parton level: top pairs}
\label{sec:tops}

%%%%%%%%%%%%%%%%%%%%%%%%%%%%%%%%%%%%%%%%%%%%%%%%%%%
\subsection{Data}
\label{sec:tops_dataset}

As a second
benchmark we apply ML-methods to top quark pair production,
unfolding from reco-level to parton level, \ie the level of the top quarks and their decay products from the hard scattering, before undergoing hadronization. While more physics assumptions/approximations are required for this type of unfolding, it is often performed by ATLAS and CMS~\cite{ATLAS:2022mlu,ATLAS:2020ccu,ATLAS:2019hxz,CMS:2024ybg,CMS:2021vhb,CMS:2020tvq,CMS:2019nrx, CMS:2019esx}.
Parton-level results are extremely useful, for instance, 
to combine measurements into a global analysis~\cite{Brivio:2019ius,Elmer:2023wtr},
extract SM parameters~\cite{Garzelli:2023rvx, CMS:2019esx},
or to compare new theory predictions without requiring these to be matched to parton showering programs~\cite{Czakon:2020qbd}.

The 
task is to map reco-level 4-momenta to parton-level 4-momenta 
defined by the $2 \to 2$ scattering and subsequent decays, in our case~\cite{Shmakov:2023kjj}
\begin{align}
    q \bar{q}/gg 
    \to t\bar{t} 
    \to (b \ell^+ \nu_\ell ) \; (\bar{b} qq) 
    \qquad \text{with} \quad \ell = e,\mu \quad , \quad q = u,d,s,c \; ,
\end{align}
plus the charge-conjugated process. The events are simulated
with Madgraph5~3.4.2~\cite{mg5} at $\sqrt{s}=13$~TeV 
and with a top quark mass $m_t=173$~GeV. One of the $W$ bosons decays leptonically,
the other hadronically. 
Showering and hadronization are simulated with Pythia~8.306~\cite{Sjostrand:2014zea}, and detector 
effects with Delphes~3.5.0~\cite{deFavereau:2013fsa} with the standard CMS card. We 
again reconstruct jets using the anti-$k_T$
algorithm~\cite{Cacciari:2008gp}, now with $R=0.5$ and a $p_T$-dependent b-tagging efficiency. 
Leptons and jets are subject to the acceptance cuts
$p_T>25$~GeV and $|\eta|<0.25$. We only keep events with exactly one
lepton, at least 2 $b$-tagged jets, and at least two more jets.

This second benchmark process is technically more challenging than the $Z$+jets unfolding
in terms of the six subjet observables,
because 
of the higher phase space dimensionality and 
because we can no longer directly match 
reco-level and parton-level observables. 
To reconstruct the hard scattering, 
the network has to learn the non-trivial combinatorics
as well as complex correlations reflected in the 
intermediate mass peaks. We  
focus on a comparison of the different 
generative unfolding methods, which reproduce the
forward simulation in their event-wise migration, but 
are most challenging from an ML-perspective. As before, we 
postpone the important question of model dependence to a later paper. 

%%%%%%%%%%%%%%%%%%%%%%%%%%%%%%%%%%%%%%%%%%%%%%%%%%%
\subsection{Generative unfolding}
\label{sec:tops_gen}

As a first attempt, we employ a straightforward 
phase space parametrization for the six
top decay products,
\begin{align}
& (p_{T, b_\ell}, \eta_{b_\ell}, \phi_{b_\ell}, \; 
   p_{T, \ell}, \eta_\ell, \phi_\ell, \; 
   p_{T, \nu}, \eta_{\nu}, \phi_{\nu}) \notag \\
& (p_{T, b_h}, \eta_{b_h}, \phi_{b_h}, m_{q_1}, \; 
   p_{T, q_1}, \eta_{q_1}, \phi_{q_1}, \; 
   p_{T, q_2}, \eta_{q_2}, \phi_{q_2})  \; .
\label{eq:naive_parametrization}
\end{align}
The lepton masses are common to all events, and we 
set them to zero at the level of our simulations.
The bottom jets are generated with a 
common finite bottom mass.
For the remaining jets, we have to keep track of 
the charm mass in the corresponding charm jets. This leads to 
a binary degree of
freedom, in addition to the 18 standard phase space dimensions. 

While at parton level all events have the same number of particles, 
at reco level we see a variable number of jets. Jets are produced in 
top and $W$-decays, but also in initial-state and final-state radiation, 
multi-parton interactions, underlying event, or pileup. Their number also strongly 
depends on the acceptance 
cuts. Naively, these additional jets are not expected to carry information on the hard 
process. However, they can sometimes cause events to pass selections by replacing top decay jets which do not pass the acceptance
cuts, or lead to challenging event reconstruction due to jet combinatorics~\cite{Heimel:2023mvw}. This means
we cannot just ignore them.

%---------------------------------------------
\begin{figure}[!t]
    \includegraphics[width=0.33\textwidth, page=17]{figs/ttbar/ttbar_naive_withVLD.pdf}
    \includegraphics[width=0.33\textwidth, page=1]{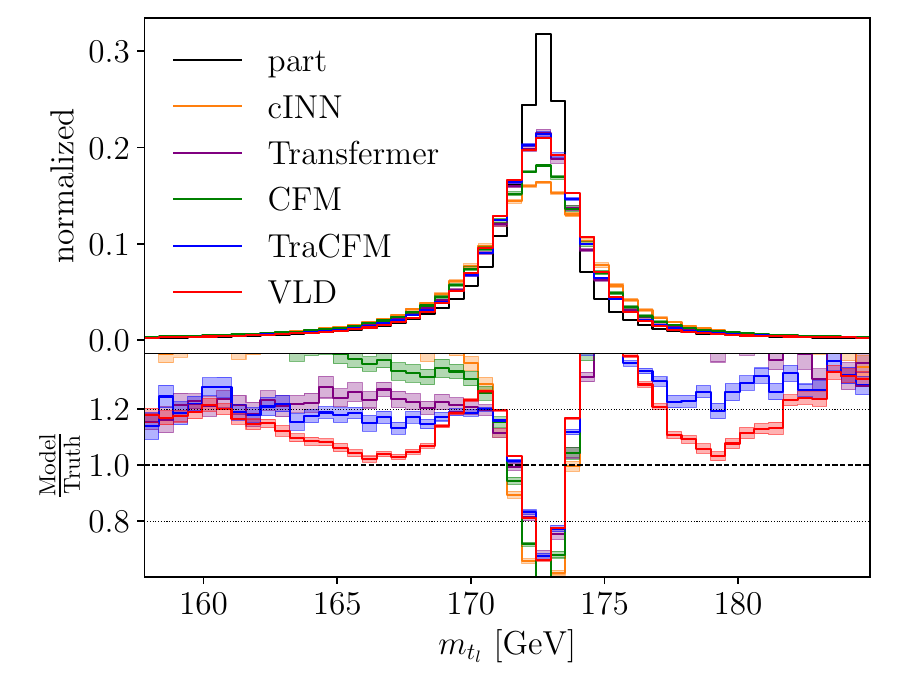}
    \includegraphics[width=0.33\textwidth, page=10]{figs/ttbar/ttbar_naive_withVLD.pdf} \\
    \includegraphics[width=0.33\textwidth, page=57]{figs/ttbar/ttbar_naive_withVLD.pdf}
    \includegraphics[width=0.33\textwidth, page=41]{figs/ttbar/ttbar_naive_withVLD.pdf}
    \includegraphics[width=0.33\textwidth, page=50]{figs/ttbar/ttbar_naive_withVLD.pdf} \\
    \includegraphics[width=0.33\textwidth, page=93]{figs/ttbar/ttbar_naive_withVLD.pdf}
    \includegraphics[width=0.33\textwidth, page=108]{figs/ttbar/ttbar_naive_withVLD.pdf}
    \includegraphics[width=0.33\textwidth, page=78]{figs/ttbar/ttbar_naive_withVLD.pdf} 
    \caption{Unfolded top pair distributions from conditional generation using the 
      naive phase space parametrization of Eq.\eqref{eq:naive_parametrization}. For the Bayesian cINN, Transfermer, CFM and TraCFM the error bars
      are based on drawing 50 samples. For the VLD the error bars are given by 
      unfolding each event 128 times and showing the bin-wise mean and standard deviation.}
    \label{fig:ttbar_results_naive}
\end{figure}
%---------------------------------------------

While the standard cINN and CFM require a fixed-dimensionality condition, 
their trans\-former variants can handle variable dimensionality. 
Alternatively, we could employ an embedding network to 
overcome this limitation. Testing the impact of additional jets on our specific 
unfolding task with the Transfermer and TraCFM networks, we find that they do 
not benefit from additional reco-level jets significantly. Consequently, we restrict 
ourselves to a fixed maximum number of particles at reco-level for these networks. The
particles we include in an ordered vector are the lepton, 
the missing transverse momentum, the two leading $b$-jets, and the two leading 
light-flavor jets.
VLD does naturally includes all jets in an un-ordered fashion.
The masses and transverse momenta of the particles are log-scaled 
before feeding them to the network.

We again use an RQS-cINN implementation from the FrEIA library~\cite{freia}, in the 
last block we replace the linear layers with Bayesian layers to track the network 
uncertainties. The CFM encodes the velocity in a standard MLP network. The time $t$ is 
embedded to a higher dimension using a random Fourier feature 
encoding~\cite{von-platen-etal-2022-diffusers} before being concatenated to the other 
network inputs, as we found that this improves results in higher-dimensional tasks. 
Following~\cite{Ernst:2023qvn} we only make the last network layer Bayesian, as too many 
Bayesian weights can make the training of large networks unstable. For the two 
Transformer-based networks we employ the standard PyTorch Transformer implementation. 
The attention blocks are then followed by a single Bayesian RQS block or a single 
Bayesian linear layer for the Transfermer and the TraCFM respectively. For the TraCFM we 
employ the same time embedding as for the CFM and concatenate the encoded time to the transformer output before feeding it to the final layer.

The results for the cINN, its Transfermer variant, the CFM, and 
its TraCFM variant along with VLD are shown in Fig.~\ref{fig:ttbar_results_naive}. 
We have checked that all generative networks reproduce the kinematics 
of the top decay products at the percent level. Generally, we
find that the lepton and neutrino kinematics are learned slightly
better than the quark kinematics. As shown in the 
top rows, the correlations describing the 
intermediate particles are not learned as well.
For the resonances, all networks struggle. Because 
they only require to correlate two independent 4-momenta,
the $W$-peaks are learned a little better than the top peaks.
Also, the leptonic decay is learned better than the hadronic decay. 
Altogether, the Transformer-enhanced networks perform better than the CFM, 
which in turn beats the cINN.

%%%%%%%%%%%%%%%%%%%%%%%%%%%%%%%%%%%%%%%%%%%%%%%%%%%
\subsection{Generative unfolding using physics}
\label{sec:tops_gen2}

%---------------------------------------------
\begin{figure}[!b]
    \includegraphics[width=0.33\textwidth, page=17]{figs/ttbar/ttbar_massparam_final.pdf}
    \includegraphics[width=0.33\textwidth, page=1]{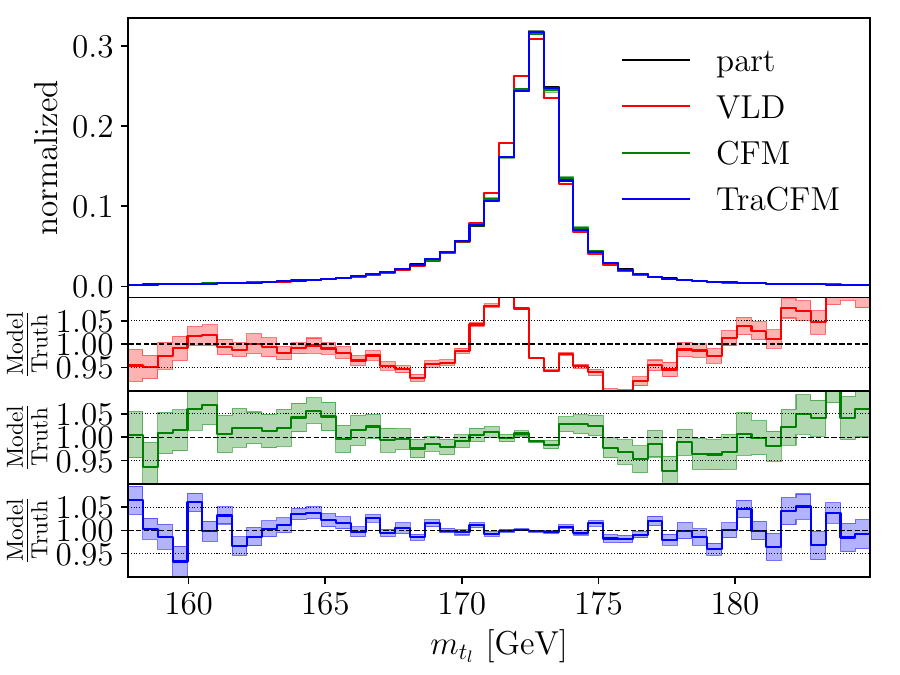}
    \includegraphics[width=0.33\textwidth, page=10]{figs/ttbar/ttbar_massparam_final.pdf} \\
    \includegraphics[width=0.33\textwidth, page=57]{figs/ttbar/ttbar_massparam_final.pdf}
    \includegraphics[width=0.33\textwidth, page=41]{figs/ttbar/ttbar_massparam_final.pdf}
    \includegraphics[width=0.33\textwidth, page=58]{figs/ttbar/ttbar_massparam_final.pdf} \\
    \includegraphics[width=0.33\textwidth, page=93]{figs/ttbar/ttbar_massparam_final.pdf}
    \includegraphics[width=0.33\textwidth, page=108]{figs/ttbar/ttbar_massparam_final.pdf}
    \includegraphics[width=0.33\textwidth, page=78]{figs/ttbar/ttbar_massparam_final.pdf}
    \caption{Unfolded top pair distributions from conditional generation using the 
      dedicated phase space parametrization of Eq.\eqref{eq:mass_parametrization}. For the Bayesian cINN, Transfermer, CFM and TraCFM the error bars
      are based on drawing 50 amples. For the VLD the error bars are given by 
      unfolding each event 32 times and showing the bin-wise mean and standard deviation.}
    \label{fig:ttbar_results_massparam}
\end{figure}
%---------------------------------------------

To solve the problems with 
intermediate on-shell propagators, we employ the dedicated
top-mass 
parametrization proposed in Ref.~\cite{Ackerschott:2023nax}. It
directly predicts the top and $W$-kinematics and makes the 
simpler decay kinematics accessible via
correlations. 
As the phase space basis we 
choose the top 4-momentum in the lab
frame, three components of the $W$ 4-momentum in the top rest frame, and
two (three for the hadronic case) components of the first $W$-decay product 
in the $W$ rest frame,
\begin{align}
&(m_t, p_{T, t}^L, \eta_t^L, \phi_t^L, \; 
  m_W, \eta_W^T, \phi_W^T, \; 
  \eta_\ell^W, \phi_\ell^W ) \notag \\
&(m_t, p_{T, t}^L, \eta_t^L, \phi_t^L, \; 
  m_W, \eta_W^T, \phi_W^T, \; 
  m_{q_1}, \eta_{q_1}^W, \phi_{q_1}^W ) \; .
\label{eq:mass_parametrization}
\end{align}
The superscripts $L,T,W$ denote the rest
frames.
We then employ a Breit-Wigner mapping using the mass values in the event generator
\begin{align}
\sqrt{2} * \text{erfinv} \left[ \frac{2}{\pi} \arctan(m-m_\text{peak}) \right] \; ,
\label{eq:breit_wigner}
\end{align}
to turn the sharp mass peaks into a Gaussian-like shape.

The results with this paramerization are shown in
Fig.~\ref{fig:ttbar_results_massparam}. We drop the cINN and 
focus on the better CFM implementations. Now that 
the intermediate masses are directly predicted by the networks, we
reproduce them within a few percent. The kinematics
of the decay particles, now correlations between the
directly predicted dimensions, are also faithfully modeled. Because the 
learning task has become easier, the 
difference between the CFM and the TraCFM is smaller. So physics 
helps, as it tends to. 

%---------------------------------------------
\begin{figure}[t]
    \centering
    \includegraphics[width=0.5\textwidth]{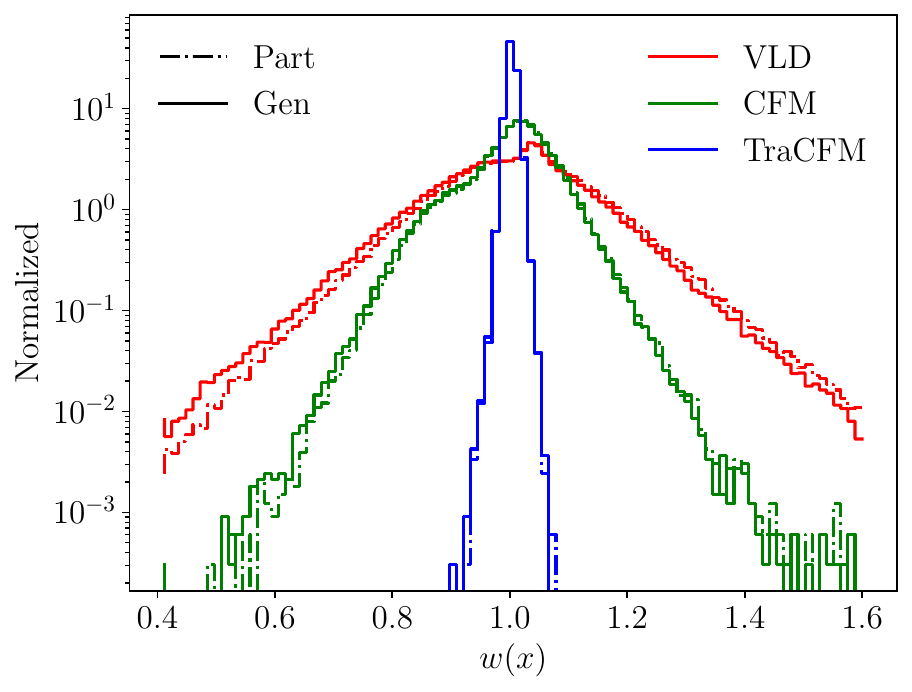}
    \caption{Weight distributions from a trained classifer between true and generated top pair events. The corresponding AUC values are 0.53 for the VLD, 0.51 for the CFM and 0.501 for the TraCFM.}
    \label{fig:ttbar_classifier}
\end{figure}
%---------------------------------------------

Similar to Sec.~\ref{sec:omni_mig}, we again train 
a classifier to distinguish the generated events from the training data truth.
Following Eq.\eqref{eq:likelihood_ratio} this classifier can be turned into a 
re-weighting function $w(x)$, evaluated for each data point in phase space.
In Fig.~\ref{fig:ttbar_classifier} we show the distribution of learned
classifier weights for the three generative unfoldings. In this case, 
we see that while the one-dimensional kinematic distributions look 
similarly good in Fig.~\ref{fig:ttbar_results_massparam}, there a
are significant differences in the precision with which the generative 
network reproduce the multi-dimensional target distributions. 
The fact that all weight distributions are peaked around $w \approx 1$ and
that the tails on the parton-level training and generated datasets are 
identical indicates that there is no definite failure mode~\cite{Das:2023ktd}.
On the other hand, the level of agreement is significantly improved,
going from the VLD to the CFM, and then adding the transformer
feature of the TraCFM to encode combinatorics~\cite{Heimel:2023mvw}. For the 
latter, we again reach the percent-level precision we got used to for the 
$Z$+jets detector unfolding in Sec.~\ref{sec:omni}.

\clearpage
%%%%%%%%%%%%%%%%%%%%%%%%%%%%%%%%%%%%%%%%%%%%%%%%%%%
\section{Outlook}
\label{sec:outlook}

Machine learning is changing the face of LHC physics, and one of the most
exciting developments is that it enables unbinned, high-dimensional,
precise unfolding. This includes detector unfolding as well as 
inverting the first-principle simulations to the parton level. There
exist three different ML-methods and tools for such an unfolding, 
(i) event reweighting or OmniFold, (ii) mapping distributions, and 
(iii) conditional generative unfolding. All these methods have been 
developed to a level, where they are ready to be further studied for use by the LHC experiments. In this paper, we give an overview of the
different methods and corresponding tools, including an update to the 
most recent neural network architectures and a rough comparison of the 
strengths of the different methods.

Our first task is to unfold detector effects for a set of six 
subjet observables in $Z$+jets production. Here, 
reweighting-based unfolding, a supervised
classification task, reproduces all true particle-level 
distributions and defines a precision benchmark shown in 
Fig.~\ref{fig:omnifold-PtoP-observables}. A new Bayesian variant of OmniFold might provide complementary strengths to the existing method.

Alternatively, 
distribution mapping can be trained on matched events efficiently. 
We found that the (Bayesian) Schr\"odinger Bridge and Direct Diffusion 
implementations
consistently provide high performance, shown 
in Fig.~\ref{fig:large_maxruntime-observables-SB_DiDi-bay}. 
Alternatively, distribution mapping can be trained on unmatched data, 
which limits its ability to reproduce the actual detector effects, but
can be useful when one is missing matched training data. 

Third, unfolding by learning and sampling conditional inverse probabilities 
is ideally suited to model complex detector
effects, but also the most challenging network architecture.
We have compared a series of tools, including invertible networks 
without and with a transformer encoding, as well as diffusion networks without and 
with a transformer, and with an enhanced latent representation. In Fig.~\ref{fig:large_maxruntime-observables-CFM_INN-bay} we have shown 
that the conditional generative tools match the precision of 
distribution mapping. In addition, we have shown to which level the
different methods learn the event migration or optimal transport 
defined by the forward detector simulation, rather than an abstract 
mapping defined by the network architecture.

Finally, we have applied our unfolding methods and tools 
to invert $t\bar{t}$ events to the hard process of top
pair production with subsequent decays.
Here, correlations pose a serious challenge, specifically 
the intermediate mass peaks. We have found that they can be learned
precisely once we represent the phase space in a physics-inspired 
kinematic basis, as can be seen in 
Fig.~\ref{fig:ttbar_results_massparam}. In addition 
to the physics pre-processing, the 
combination of a diffusion model with a transformer 
guaranteed the best performance among the conditional 
generative unfolding networks.

Altogether, we have shown a 
multitude of different methods and tools for ML-
unfolding, with dedicated individual strengths.\footnote{Many of the codes used in this paper will be made \href{https://github.com/heidelberg-hepml/ml-tutorials}{publicly available}, together with a set of tutorials accompanying Ref.~\cite{Plehn:2022ftl}.} All of 
them are ready to be studied further in the context of LHC 
analyses. Their complementarity is a strength for building 
confidence in advanced tools for high-dimensional cross 
section measurements.

%%%%%%%%%%%%%%%%%%%%%%%%%%%%%%%%%%%%%%%%%%%%%%%%%%%
\clearpage
\subsection*{Acknowledgements}

We would like to thank Sofia Palacios Schweitzer for her crucial contributions to 
DiDi-unfolding and Bogdan Malaescu for many enlightening discussions on 
unfolding.
TP and AB would like to thank the Baden-W\"urttemberg-Stiftung for financing through the program \textit{Internationale Spitzenforschung}, project
\textsl{Uncertainties --- Teaching AI its Limits}
(BWST\_IF2020-010). The Heidelberg group is supported by 
the KISS consortium (05D23GU4) funded by BMBF 
in the ErUM-Data action plan,
the Deutsche
Forschungsgemeinschaft (DFG, German Research Foundation) under grant
396021762 -- TRR~257 \textsl{Particle Physics Phenomenology after the
Higgs Discovery}, and through Germany's
Excellence Strategy EXC~2181/1 -- 390900948 (the \textsl{Heidelberg
  STRUCTURES Excellence Cluster}).  SD, VM, and BN are supported by the U.S. Department of Energy (DOE), Office of Science under contract DE-AC02-05CH11231.
  DW, KG, and MF are supported by DOE grant DE-SC0009920. This research used resources of the National Energy Research Scientific Computing Center, a DOE Office of Science User Facility supported by the Office of Science of the U.S. Department of Energy under Contract No. DE-AC02-05CH11231 using NERSC award HEP-ERCAP0021099.  
  
\clearpage
\appendix
%%%%%%%%%%%%%%%%%%%%%%%%%%%%%%%%%%%%%%%%%%%%%%%%%%
\section{Combined $Z$+jets results}
\label{sec:combined}

\begin{figure}[H]
    \includegraphics[width=0.495\textwidth, page=1]{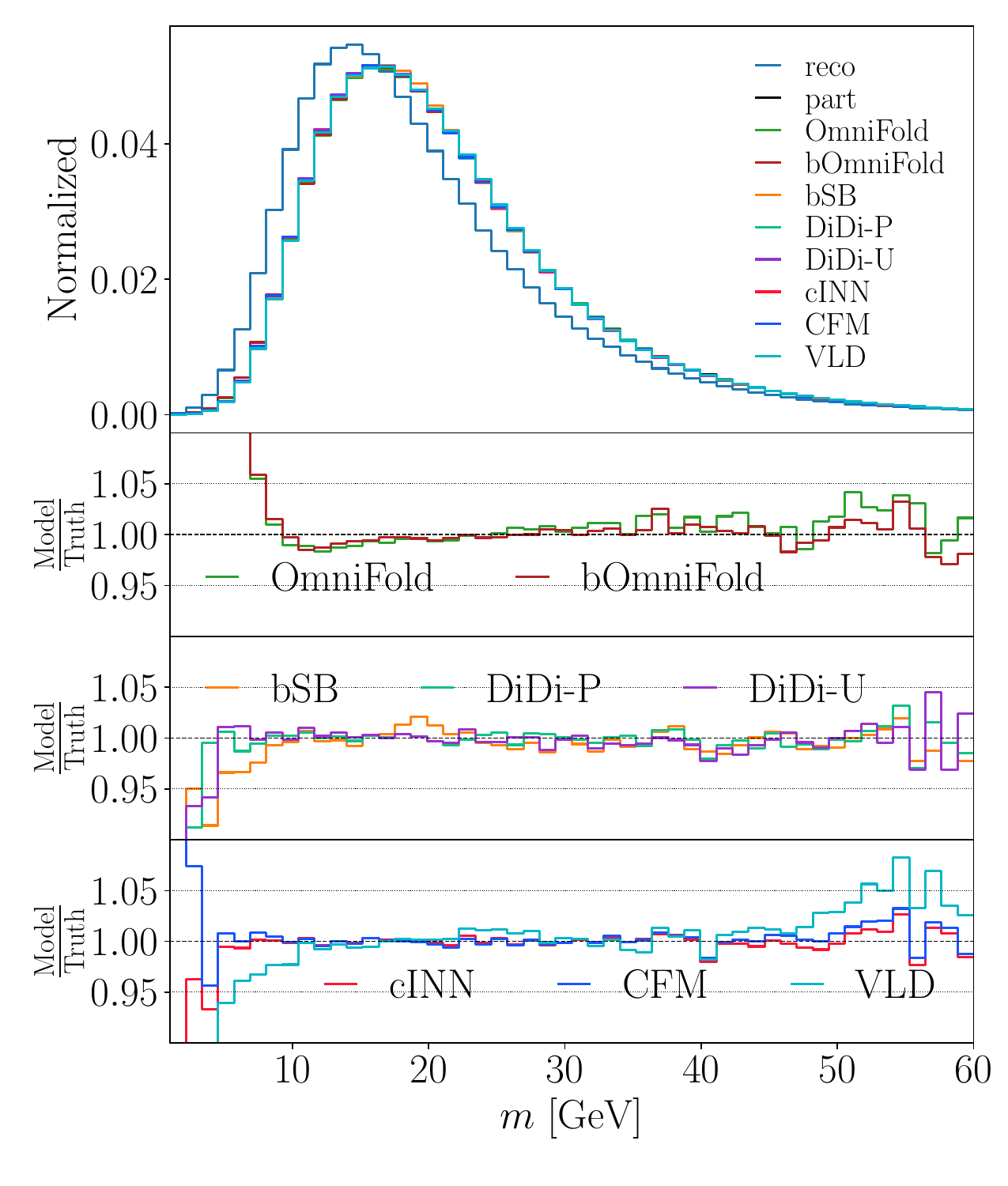}
    \includegraphics[width=0.495\textwidth, page=2]{figs/new_dataset/summary_plot.pdf}\\
    \includegraphics[width=0.495\textwidth, page=3]{figs/new_dataset/summary_plot.pdf} 
    \includegraphics[width=0.495\textwidth, page=6]{figs/new_dataset/summary_plot.pdf}
    \caption{Results collected from Sec.~\ref{sec:omni}, showing all unfolding networks, as well as the (b)Omnifold de-noising benchmark.}
    \label{fig:summary_plot}
\end{figure}

In Fig.~\ref{fig:summary_plot} we compare the unfolding results for 
$Z$+jets events, as discussed in Sec.~\ref{sec:omni}. We show
the same kinematic observables as in Fig~\ref{fig:omnifold-PtoP-observables} 
for the (b)OmniFold benchmark, in 
Fig.~\ref{fig:large_maxruntime-observables-SB_DiDi-bay} for the 
distributions matching, and in Fig.~\ref{fig:large_maxruntime-observables-CFM_INN-bay} for the conditional generation. We omit all error bars representing 
statistical or Bayesian network uncertainties. The (b)OmniFold curves 
do not show unfolding, but learned densities from noisy data. 
They can be compared to each other, and 
used as benchmarks for the actual unfolding. None 
of the networks have been especially optimized for the task, so for all 
of them there should still be possible performance gains.

%------------------------------------------------------
%------------------------------------------------------

\clearpage
%%%%%%%%%%%%%%%%%%%%%%%%%%%%%%%%%%%%%%%%%%%%%%%%%%
\section{Hyperparameters}
\label{sec:hyperparams}

%----------------------------------------------------------
\begin{table}[ht!]
    \centering
    \begin{small} \begin{tabular}[t]{l|cccc}
    \toprule
    Parameter & DiDi & CFM $Z$+jets & CFM $t\bar{t}$ & TraCFM\\
    \midrule
    Optimizer & \multicolumn{4}{c}{Adam}  \\
    Learning rate & \multicolumn{4}{c}{0.001}  \\
    LR schedule & \multicolumn{4}{c}{Cosine annealing} \\
    Batch size & \multicolumn{4}{c}{16384}  \\
    Epochs & 500 & 400  & 1000  & 500 \\
    Network&  MLP & MLP & MLP  & Transformer \\
    Number of layers & 8 & 8 & 8 & -\\
    Hidden nodes & 512 & 512 & 1024 & -\\
    Transformer blocks & - & - & - & 6 \\
    Transformer heads & - & - & - & 4 \\
    Embedding dim & - & - & - & 128 \\
    Bayesian regularization & \multicolumn{4}{c}{1} \\
    \bottomrule
    \end{tabular} \end{small}
    \caption{Network and training hyperparameters for the Direct Diffusion and CFM networks in Figs.~\ref{fig:large_maxruntime-observables-SB_DiDi-bay}, \ref{fig:large_maxruntime-observables-CFM_INN-bay}, \ref{fig:ttbar_results_naive}, and~\ref{fig:ttbar_results_massparam} }
    \label{tab:didicfm_hyperparams}
\end{table}
%----------------------------------------------------------

%----------------------------------------------------------
\begin{table}[ht!]
    \centering
    \begin{small} \begin{tabular}[t]{l|ccc}
    \toprule
    Parameter & cINN $Z$+jets & cINN $t\bar{t}$ & Transfermer\\
    \midrule
    Optimizer & \multicolumn{3}{c}{Adam}  \\
    Max Learning rate & \multicolumn{3}{c}{0.0003}  \\
    LR schedule & \multicolumn{3}{c}{One cycle} \\
    Batch size & \multicolumn{3}{c}{1024}  \\
    Epochs & 75 & 130  & 250 \\
    Network&  RQS-INN & RQS-INN & Transformer+RQS \\
    INN blocks & 10 & 20 & 1 \\
    RQS bins & 24 & 30 & 30 \\
    Subnet layers & 5 & 5 & 5 \\
    Subnet dim & 200 & 256 & 256 \\
    Transformer blocks & - & - & 6  \\
    Transformer heads & - & - & 4 \\
    Embedding dim & - & - & 128  \\
    \bottomrule
    \end{tabular} \end{small}
    \caption{Network and training hyperparameters for the cINN and Transfermer in Figs.~\ref{fig:large_maxruntime-observables-CFM_INN-bay} 
    and~\ref{fig:ttbar_results_naive}.}
    \label{tab:inn_hyperparams}
\end{table}
%----------------------------------------------------------

%----------------------------------------------------------
\begin{table}[ht!]
    \centering
    \begin{small} \begin{tabular}[t]{l|c}
    \toprule
    Parameter & SB\\
    \midrule
    Optimizer & Adam\\
    Learning rate &  0.001\\
    Batch size & 128 \\
    Network Updates &  250000\\
    Network &   Fully connected ResNet \\
    Blocks & 6\\
    MLP size & 256 \\
    \bottomrule
    \end{tabular} \end{small}
    \caption{Network and training hyperparameters for the Schr\"odinger Bridge in Fig.~\ref{fig:large_maxruntime-observables-SB_DiDi-bay}.}
    \label{tab:sb_hyperparams}
\end{table}
%----------------------------------------------------------

%----------------------------------------------------------
\begin{table}[t]
    \centering
    \begin{small} \begin{tabular}[t]{l|cc}
    \toprule
    Parameter & VLD $Z$+jets & VLD $t\bar{t}$ \\
    \midrule
    Optimizer & Adam &  Adam \\
    Initial Learning rate & $5 \times 10^{-4}$ & $5 \times 10^{-4}$ \\
    Fine-tune Learning rate & $1 \times 10^{-4}$ & $1 \times 10^{-4}$ \\
    Batch size & $1024$ & $1024$ \\
    Updates & $1$ Million & $1$ Million \\
    Hidden Dimensions & 64 & 64  \\
    Denoising Layers & 8 & 8\\
    Detector Encoder Layers & 6 & 6\\
    Part* Encoder Layers & 6 & 6\\
    Part* Decoder Layers & 6 & 6\\
    \bottomrule
    \end{tabular} \end{small}
    \caption{Network and training hyperparameters for the VLD networks in Figs~\ref{fig:large_maxruntime-observables-CFM_INN-bay}, \ref{fig:ttbar_results_naive}, and~\ref{fig:ttbar_results_massparam}.}
    \label{tab:vld_hyperparams}
\end{table}
%----------------------------------------------------------

%%%%%%%%%%%%%%%%%%%%%%%%%%%%%%%%%%%%%%%%%%%%%%%%%%
\phantomsection\bibliographystyle{tepml}
\bibliography{ben,daniel,tilman,generative,refs,daniel_old,ben_old}
\end{document}

%% file: incl_settings.tex
\usepackage[utf8]{inputenc} % input encodings (allow UTF-8 input)
\usepackage[T1]{fontenc} 	% selecting font encodings (use 8-bit T1 fonts)
\usepackage[english]{babel} % multilingual support (English language/hyphenation)

\usepackage[bitstream-charter]{mathdesign}
\usepackage{geometry} 		% flexible and complete interface to document dimensions
\usepackage{amsmath} 		% math package (American Mathematical Society)
\usepackage{mathtools} 		% math package (fixes various deficiencies of amsmath)
\usepackage{float} 			% floating objects such as figures and tables
\usepackage{graphicx} 		% enhanced support for graphics
\usepackage{tabularx} 		% tabulars with adjustable-width columns
\usepackage{booktabs} 		% professional-quality tables
\usepackage{color, xcolor} 	% foreground and background colour management
\usepackage{pdfpages} 		% inclusion of external multi-page PDF documents
\usepackage{extarrows} 		% extra arrows beyond those provided in amsmath
\usepackage{multirow} 		% create tabular cells spanning multiple rows
\usepackage{multicol} 		% intermix single and multiple columns
\usepackage{enumitem} 		% control layout of itemize, enumerate, description
\usepackage{xspace} 		% define commands that appear not to eat spaces
\usepackage{stackrel} 		% enhancement to the \stackrel command
\usepackage{tikz} 			% create PostScript and PDF graphics
\usepackage{braket} 		% Dirac bra-ket notation
\usepackage{bm} 			% access bold symbols in maths mode
\usepackage{tensor} 		% typeset tensors (tensor-style super- and subscripts)
\usepackage{slashed} 		% slash through characters (Feynman slash notation)
\usepackage{siunitx} 		% SI units package (typesetting values with units)
\usepackage{lastpage} 		% reference last page
\usepackage{cite} 			% improved citation handling
\usepackage[normalem]{ulem} % package for underlining
\usepackage{fontawesome} 	% access to web-related icons
\usepackage{tocloft} 		% control over the typography of the TOC, LOF, and LOT
\usepackage{titlesec} 		% interface to sectioning commands (various title styles)
\usepackage{doi} 			% create correct hyperlinks for DOI numbers
\usepackage{hyperref} 		% hypertext links (handel cross-referencing commands)
\usepackage[most]{tcolorbox} 					% coloured and framed text boxes
\usepackage[nameinlink, capitalize]{cleveref} 	% intelligent cross-referencing
\usepackage[nottoc, notlot, notlof]{tocbibind} 	% add references/index/contents to TOC
\usepackage[ruled, vlined]{algorithm2e} 		% floating algorithm environment
\usepackage{makecell}
\usepackage[makeroom]{cancel}
\usepackage{feynmf}

\makeatletter
\def\BState{\State\hskip-\ALG@thistlm}
\makeatother

\makeatletter
\@ifundefined{pdfoutput}{}{\DeclareGraphicsRule{*}{mps}{*}{}}
\makeatother

\makeatletter
\DeclareRobustCommand*{\bfseries}{%
   \not@math@alphabet\bfseries\mathbf
   \fontseries\bfdefault\selectfont
   \boldmath
}
\makeatother

\hypersetup{
	pdftitle={Modern Machine Learning Tools for Unfolding},
	pdfauthor={Huetsch et al.},
	colorlinks=true, 			% false: boxed links, true: colored links
	linkcolor={red!50!black}, 	% color of internal links (sections, pages, etc.)
	citecolor={blue!50!black}, 	% color of citation links (links to bibliography)
	urlcolor={blue!80!black} 	% color of URL links (external links)
} 
\DeclareSymbolFont{usualmathcal}{OMS}{cmsy}{m}{n}
\DeclareSymbolFontAlphabet{\mathcal}{usualmathcal}

\SetArgSty{textnormal}
\SetKwComment{Comment}{{\small\#}~}{}
\SetCommentSty{mycommfont}

\setitemize{itemsep=2pt, parsep=0pt} 				% adjust itemize environment
\setenumerate{itemsep=2pt, parsep=0pt} 				% adjust enumerate environment
\crefname{section}{Sec.}{Secs.}
\crefname{equation}{Eq.}{Eqs.}
\crefname{figure}{Fig.}{Figs.}
\crefname{table}{Tab.}{Tabs.}
\Crefname{section}{Section}{Sections}
\Crefname{equation}{Equation}{Equations}
\Crefname{figure}{Figure}{Figures}
\Crefname{table}{Table}{Tables}

\usetikzlibrary{arrows, shapes.geometric, arrows.meta, shapes, decorations.pathreplacing, fit, patterns, patterns.meta}

\definecolor{Rcolor}{HTML}{E99595}
\definecolor{Gcolor}{HTML}{C5E0B4}
\definecolor{Bcolor}{HTML}{9DC3E6}
\definecolor{Ycolor}{HTML}{FFE699}

\tikzstyle{expr} = [circle, minimum width=1.8cm, minimum height=1.8cm, text centered, align=center, inner sep=0, draw,font=\LARGE]
\tikzstyle{txt_huge} = [align=center, font=\Huge, scale=2]
\tikzstyle{txt} = [align=center, font=\LARGE]
\tikzstyle{cinn} = [double arrow, double arrow head extend=0cm, double arrow tip angle=130, shape border rotate=90, inner sep=0, align=center, minimum width=2.1cm, minimum height=2.3cm, fill=Bcolor, draw,font=\LARGE]
\tikzstyle{cinn_black} = [cinn, minimum height=2.5cm, fill=black]
\tikzstyle{arrow} = [thick,-{Latex[scale=1.0]}, line width=0.2mm, color=black]
\tikzstyle{line} = [thick, line width=0.2mm, color=black]
\tikzstyle{loss} = [rectangle, align=center,  minimum width=1.8cm, minimum height=1.5cm,fill=Rcolor,font=\LARGE, rounded corners]
\tikzstyle{xt} = [rectangle, align=center,  minimum width=4cm, minimum height=1.5cm,fill=Gcolor,font=\Large, rounded corners]
\tikzstyle{xts} = [rectangle, align=center,  minimum width=1cm, minimum height=1.5cm,fill=Gcolor,font=\Large, rounded corners]

%% file: incl_shortcuts.tex
\definecolor{red_cb}{HTML}{e41a1c}
\definecolor{blue_cb}{HTML}{377eb8}
\definecolor{green_cb}{HTML}{4daf4a}
\definecolor{purple_cb}{HTML}{984ea3}
\definecolor{orange_cb}{HTML}{ff7f00}

\definecolor{EmeraldGreen}{HTML}{1ea78d}
\definecolor{EnglishRed}{HTML}{b02427}
\newcommand{\ie}{\text{i.e.}\;}

\newcommand{\kl}{\text{KL}}
\newcommand{\pl}{p_\text{latent}}
\newcommand{\pd}{p_\text{data}}
\newcommand{\psimr}{p_\text{sim}}
\newcommand{\psimp}{p_\text{gen}}
\newcommand{\pmd}{p_\text{model}}
\newcommand{\punf}{p_\text{unfold}}

\newcommand{\xp}{x_\text{part}}
\newcommand{\xr}{x_\text{reco}}

\newcommand{\XLangle}{\Bigl\langle}
\newcommand{\XRangle}{\Bigr\rangle}
\newcommand{\XXLangle}{\biggl\langle}
\newcommand{\XXRangle}{\biggr\rangle}

\newcommand{\qqquad}{\qquad\quad}

\newcommand\one{\leavevmode\hbox{\small1\normalsize\kern-.33em1}}
\newcommand{\norm}[1]{\lVert#1\rVert} 		% norm (double vertical lines)
\newcommand{\normal}{\mathcal{N}}

\newcommand{\loss}{\mathcal{L}} 	% loss value

\newcommand{\arXiv}[2][]{%
	\ifthenelse{\equal{#1}{}}%
	{\href{http://arxiv.org/abs/#2}{arXiv:#2}}%
	{\href{http://arxiv.org/abs/#2}{arXiv:#2~[#1]}}}

\def\slashchar#1{\setbox0=\hbox{$#1$}           % set a box for #1
   \dimen0=\wd0                                 % and get its size
   \setbox1=\hbox{/} \dimen1=\wd1               % get size of /
   \ifdim\dimen0>\dimen1                        % #1 is bigger
      \rlap{\hbox to \dimen0{\hfil/\hfil}}      % so center / in box
      #1                                        % and print #1
   \else                                        % / is bigger
      \rlap{\hbox to \dimen1{\hfil$#1$\hfil}}   % so center #1
      /                                         % and print /
   \fi}

\newcommand{\tikznode}[2]{%
\ifmmode%
\tikz[remember picture,baseline=(#1.base),inner sep=0pt] \node (#1) {$#2$};%
\else
\tikz[remember picture,baseline=(#1.base),inner sep=0pt] \node (#1) {#2};%
\fi}

\def\mathswitchr#1{\relax\ifmmode{\mathrm{#1}}\else$\mathrm{#1}$\xspace\fi}
\def\mathswitch#1{\relax\ifmmode#1\else$#1$\xspace\fi}

\DeclarePairedDelimiterX{\infdivx}[2]{[}{]}{%
  #1\;\delimsize\|\;#2%
}

%% file: tra_cfm.tex
\definecolor{Rcolor}{HTML}{E99595}
\definecolor{Gcolor}{HTML}{C5E0B4}
\definecolor{Gcolor_light}{HTML}{F1F8ED}
\definecolor{Bcolor}{HTML}{9DC3E6}
\definecolor{Ycolor}{HTML}{FFE699}
\definecolor{Ycolor_light}{HTML}{FFF7DE}

\tikzstyle{expr} = [rectangle, rounded corners=0.3ex, minimum width=1.5cm, minimum height=1cm, text centered, align=center, inner sep=0, fill=Ycolor, font=\large, draw]

\tikzstyle{small_cinn} = [double arrow, double arrow head extend=0cm, double arrow tip angle=130, inner sep=0, align=center, minimum width=1.1cm, minimum height=0.5cm, fill=Rcolor, draw]

\tikzstyle{small_cinn_black} = [small_cinn, minimum height=1.5cm, fill=black]

\tikzstyle{transformer} = [rectangle, rounded corners, minimum width=6cm, minimum height=2.4cm, font=\large, fill=Gcolor_light, draw]

\tikzstyle{attention} = [rectangle, rounded corners=0.3ex, minimum width=5.5cm, minimum height=1.2cm, align=center, fill=Gcolor, draw, font=\large]

\tikzstyle{txt_huge} = [align=center, font=\large, scale=2]
\tikzstyle{txt} = [align=center, font=\large, minimum height=1cm]

\tikzstyle{arrow} = [thick,-{Latex[scale=1.0]}, line width=0.2mm, color=black]
\tikzstyle{line} = [thick, line width=0.2mm, color=black]

\begin{tikzpicture}[node distance=2cm, scale=0.75, every node/.style={transform shape}]

\node (part1) [txt] {$x_{\text{reco}}^{(1)}$};
\node (part2) [txt, right of=part1, xshift=-0.5cm] {$...$};
\node (part3) [txt, right of=part2, xshift=-0.5cm] {$x_{\text{reco}}^{(n_r)}$};

\node (emb_part1) [expr, below of=part1, yshift=-0.3cm, rotate=90]{Emb};
\node (emb_part2) [txt, below of=part2, yshift=-0.3cm] {$...$};
\node (emb_part3) [expr, below of=part3,yshift=-0.3cm,  rotate=90]{Emb};

\node (TE) [transformer, below of=emb_part2, yshift=-0.7cm, text width=4cm,
text depth=1.5cm, align=center] {Transformer-Encoder};
\node (TE_att) [attention, below of=TE, yshift=1.6cm] {Self-Attention \\
Reco-level correlations};

\node (reco1) [txt, right of=part3, xshift=1.75cm] {$x_{\text{part}}^{(1)}(t)$};
\node (reco3) [txt, right of=reco1] {$...$};
\node (reco6) [txt, right of=reco3] {$x_{\text{part}}^{(n_p)}(t)$};

\node (t) [txt, right of=reco6, xshift=0.5cm] {$t$};

\node (emb_reco1) [expr, below of=reco1, yshift=-0.3cm, rotate=90]{Emb};
\node (emb_reco3) [txt, below of=reco3, yshift=-0.3cm] {$...$};
\node (emb_reco6) [expr, below of=reco6, yshift=-0.3cm, rotate=90]{Emb};

\node (TD) [transformer, right of=TE, xshift=5.3cm, yshift=-0.8cm, text width=4cm,
text depth=3.1cm, align=center, minimum height=4cm] {Transformer-Decoder};
\node (TD_att) [attention, right of=TE_att, xshift=5.3cm] {Self-Attention \\
Part-level correlations};
\node (TD_crossatt) [attention, below of=TD_att, yshift=0.4cm] {Cross-Attention \\
Combinatorics};

%\node (inn1_b) [small_cinn_black, below of=reco1, yshift=-9.5cm, rotate=90]{cINN};
%\node (inn6_b) [small_cinn_black, below of=reco6, yshift=-9.5cm, rotate=90]{cINN};
\node (inn1) [small_cinn, below of=reco1, yshift=-8.2cm, rotate=90]{Linear};
\node (inn3) [txt, below of=reco3, yshift=-8.2cm]{$...$};
\node (inn6) [small_cinn, below of=reco6, yshift=-8.2cm, rotate=90]{Linear};

%\node (linear) [attention, below of=TD, yshift=-2.2cm, fill=Rcolor]{Linear Layer};

\node (prob1) [txt, below of=inn1, yshift=-0.3cm]{$\Big( v^{(1)}(c^{(1)}, t ),$ };
\node (prob2) [txt, below of=inn3, yshift=-0.3cm]{$...$};
\node (prob6) [txt, below of=inn6, yshift=-0.3cm]{$, \; v^{(n)}(c^{(n_p)}, t ) \Big)$};
\node (prob) [txt, left of=prob1, xshift=-2.5 cm]{$v(x_\text{part}(t), t | x_\text{reco}) = \;$};

\draw [arrow, color=black] (part1.south) -- (emb_part1.east);
\draw [arrow, color=black] (part3.south) -- (emb_part3.east);

\draw [arrow, color=black] (emb_part1.west) -- (TE.north -| emb_part1.west);
\draw [arrow, color=black] (emb_part3.west) -- (TE.north -| emb_part3.west);

\draw [arrow, color=black] (TE.south -| emb_part1.west) --  ([yshift=-1cm]TE.south -| emb_part1.west) -- ([yshift=-1cm]TE.south -| TD.west) ; 
\draw [arrow, color=black] (TE.south -| emb_part3.west) --  ([yshift=-0.7cm]TE.south -| emb_part3.west) -- ([yshift=-0.7cm]TE.south -| TD.west);

\draw [arrow, color=black] ([xshift=-0.2cm]reco1.south) -- ([xshift=-0.2cm]emb_reco1.east);
\draw [arrow, color=black] ([xshift=-0.2cm]reco6.south) -- ([xshift=-0.2cm]emb_reco6.east);

\draw [arrow, color=black] (emb_reco1.west) -- (TD.north -| emb_reco1.west);
\draw [arrow, color=black] (emb_reco6.west) -- (TD.north -| emb_reco6.west);

(A) (B);

\draw [arrow, color=black] ([xshift=-0.2cm]TD.south -| emb_reco1.west)  -- node [text width=1.5cm, pos=0.3, font=\large, right] {$c^{(1)}$} ([xshift=-0.2cm]inn1.east -| emb_reco1.west);
\draw [arrow, color=black] ([xshift=-0.2cm]TD.south -| emb_reco6.west)  -- node [text width=1.5cm,pos=0.3, font=\large, right] {$c^{(n_p)}$}  ([xshift=-0.2cm]inn6.east -| emb_reco6.west);

\draw [arrow, color=black] (inn1.west -| emb_reco1.west) -- (prob1.north -| emb_reco1.west);
\draw [arrow, color=black] (inn6.west -| emb_reco6.west) -- (prob6.north -| emb_reco6.west);

\draw [arrow, color=black] (t.south) --  ([yshift=-0.5cm]t.south) -- ([yshift=-0.5cm, xshift=0.2cm]t.south -| reco1.center) -- ([xshift=0.2cm]emb_reco1.east); 
\draw [arrow, color=black] ([yshift=-0.5cm, xshift=0.2cm]t.south -| reco6.center) -- ([xshift=0.2cm]emb_reco6.east); 
%\draw [arrow, color=black] (t.south) --  (t.south |- linear.east) -- (linear.east) ; 
\draw [arrow, color=black] (t.south) --  ([yshift=-8.1cm]t.south) -- ([yshift=-8.1cm, xshift=0.2cm]t.south -| reco1.center) -- ([xshift=0.2cm]inn1.east); 
\draw [arrow, color=black] ([yshift=-8.1cm, xshift=0.2cm]t.south -| reco6.center) -- ([xshift=0.2cm]inn6.east); 

\end{tikzpicture}